\documentclass[graybox,footinfo]{svmult}

\usepackage{mathptmx}       
\usepackage{helvet}         
\usepackage{courier}        

\usepackage{makeidx}         
\usepackage{graphicx}        
\usepackage{multicol}        
\usepackage[bottom]{footmisc}

\usepackage{url}

\makeindex             

\begin{document}

\title*{Excitation of Nucleobases from a Computational Perspective II: Dynamics
\thanks{
This manuscript was originally published under \textit{S. Mai, M. Richter, P. Marquetand \& L. Gonz\'alez Barbatti, M., in Borin, A. C. \& Ullrich, S. (Eds.) Excitation of Nucleobases from a Computational Perspective II: Dynamics Top. Curr. Chem., Photoinduced Phenomena in Nucleic Acids I, Springer Berlin Heidelberg, 2015, 355, 99-153}.
The final publication is available at Springer via http://dx.doi.org/10.1007/128\_2014\_549.
}
}
\author{Sebastian Mai \and Martin Richter \and Philipp Marquetand \and Leticia Gonz\'alez}
\institute{Sebastian Mai \at Institute of Theoretical Chemistry, University of Vienna, W\"ahringer Str. 17, 1090 Vienna, Austria,
\and Martin Richter \at Institute of Theoretical Chemistry, University of Vienna, W\"ahringer Str. 17, 1090 Vienna, Austria,
\and Philipp Marquetand \at Institute of Theoretical Chemistry, University of Vienna, W\"ahringer Str. 17, 1090 Vienna, Austria,
\and Leticia Gonz\'alez \at Institute of Theoretical Chemistry, University of Vienna, W\"ahringer Str. 17, 1090 Vienna, Austria, \email{leticia.gonzalez@univie.ac.at}
}

\maketitle


\abstract*{
This Chapter is devoted to unravel the relaxation processes taking place after photoexcitation of isolated DNA/RNA nucleobases in gas phase from a time-dependent perspective. To this aim, several methods are at hand, ranging from full quantum dynamics to various flavours of semiclassical or ab initio molecular dynamics, each with its advantages and its limitations. As this contribution shows, the most common approach employed up-to-date to learn about the deactivation of nucleobases in gas phase is a combination of the Tully surface hopping algorithm with on-the-fly CASSCF calculations. Different methods or, even more dramatically, different electronic structure methods can provide different dynamics. A comprehensive review of the different mechanisms suggested for each nucleobase is provided and compared to available experimental time scales. The results are discussed in a general context involving the effects of the different applied electronic structure and dynamics methods. Mechanistic similarities and differences between the two groups of nucleobases---the purine derivatives (adenine and guanine) and the pyrimidine derivatives (thymine, uracil, and cytosine)---are elucidated. Finally, a perspective on the future of dynamics simulations in the context of nucleobase relaxation is given.
}


\clearpage
\section*{Contents}
\setcounter{minitocdepth}{2}
\dominitoc

\section*{Abbreviations}

\begin{description}[TD-DFTB]
  \item[A] Adenine
  \item[AIMD] Ab initio molecular dynamics
  \item[AIMS] Ab initio multiple spawning
  \item[AM1] (Semi-empirical) Austin model 1
  \item[C] Cytosine
  \item[CASPT2] Complete active space second-order perturbation theory
  \item[CASSCF] Complete active space self-consistent field
  \item[CI] Configuration interaction
  \item[CoIn] Conical intersection
  \item[CPMD] Car-Parinello molecular dynamics
  \item[cs] Closed shell
  \item[DFT] Density functional theory
  \item[DFTB] Density functional-based tight binding
  \item[DNA] Deoxyribonucleic acid
  \item[DOF] Degree of freedom
  \item[FC] Franck-Condon
  \item[FMS] Full multiple spawning
  \item[G] Guanine
  \item[GS] Ground state
  \item[IC] Internal conversion
  \item[ISC] Intersystem crossing
  \item[MCH] Molecular Coulomb Hamiltonian
  \item[MCTDH] Multi-configurational time-dependent Hartree
  \item[MD] Molecular dynamics
  \item[MRCI] Multi-reference configuration interaction
  \item[MRCIS] Multi-reference configuration interaction with single excitations
  \item[NAC] Non-adiabatic coupling
  \item[OM2] (Semi-empirical) Orthogonalization model 2
  \item[PEH] Potential energy hypersurface
  \item[PM3] (Semi-empirical) Parametrized model 3
  \item[QD] Quantum dynamics
  \item[RNA] Ribonucleic acid
  \item[ROKS] Restricted open-shell Kohn-Sham
  \item[\textsc{Sharc}] Surface hopping including arbitrary couplings
  \item[SOC] Spin-orbit coupling
  \item[T] Thymine
  \item[TD-DFT] Time-dependent density functional theory
  \item[TD-DFTB] Time-dependent density functional-based tight binding
  \item[TDSE] Time-dependent Schr\"odinger equation
  \item[TRPES] Time-resolved photo-electron spectroscopy
  \item[TSH] Tully's surface hopping
  \item[TSH-CP] Tully's surface hopping coupled to Car-Parinello dynamics
  \item[U] Uracil
  \item[UV] Ultra-violet
\end{description}


\section{Introduction}\label{sec:introduction}

A wealth of reactions can occur after a molecule is excited by electromagnetic radiation. Especially important for all life on earth is the interaction of nucleic acids with ultraviolet (UV) light. The reason is that the genetic information, which deoxyribonucleic acid (DNA) or ribonucleic acid (RNA) carry, can be damaged by photoreactions~\cite{Kleinermanns2013IRPC, Middleton2009ARPC, Markovitsi2007PPS, Shukla2007JBSD, Saigusa2006JPPC, Crespo-Hernandez2004CR} leading e.g. to skin cancer, which is the most frequent type of cancer~\cite{Leiter2008}. DNA/RNA and in particular DNA/RNA nucleobases are photostable, meaning that they have mechanisms to return to the electronic ground state soon after light irradiation, thus avoiding detrimental excited-state reactions.

The question of how photostability is achieved on an atomistic level has motivated a large amount of theoretical and experimental studies. On the microscopic scale, reactions are rearrangements of electrons and nuclei in time. Hence, the natural picture for an investigation of these processes is an approach where the instantaneous dynamics of the particles is described. This ansatz can be pursued computationally and constitutes the field of dynamics simulations~\cite{Tannor2006,Meyer2003TCA,Martinez2006ACR,Barbatti2011WCMS,Plasser2012TCA,Curchod2013CPC}. Here, a detailed picture of the mechanisms underlying a photophysical and photochemical processes is obtained in an intuitive fashion. Important pathways along essential points of the reaction, like minima, transition states or surface crossings, are naturally identified. Of course, such reaction points can be also obtained with static quantum mechanical calculations, as illustrated in the previous Chapter, and possible reaction routes can be suggested. Dynamics simulations, however, give an unequivocal answer regarding the regions which are \emph{de facto} visited by the molecule after it gets electronically excited. Moreover, dynamics calculations naturally provide time-scales and quantum yields that can be compared with time-resolved spectroscopic experiments.

\begin{figure}
  \includegraphics[scale=1]{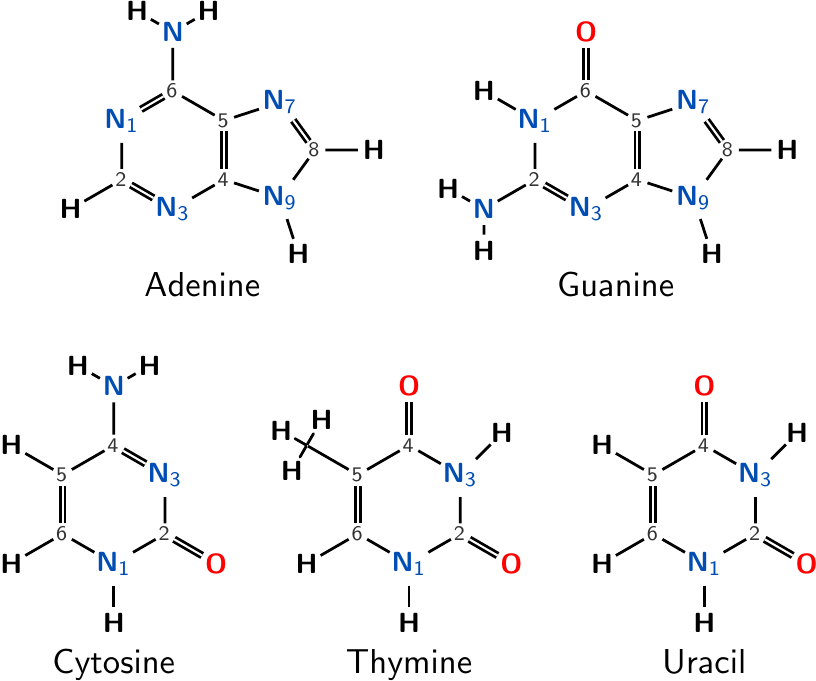}
  \caption{The chemical structures and atom indices of the five canonical nucleobases of DNA/RNA. In the top row the purine bases are given, in the bottom row the pyrimidine bases.}
  \label{fig:intro:molecules}
\end{figure}

The present chapter deals with dynamical simulations of isolated nucleobases electronically excited by UV light. Within the nucleic acids, the most important chromophores, i.e.\ the moieties which absorb the light, are the nucleobases Adenine (A), Guanine (G), Cytosine (C), Thymine (T) or Uracil (U), which pairwise constitute the bridges of the well-known double helix structure~\cite{Watson1953N}. They can be grouped into the purine bases (A and G) and the pyrimidine bases (C, T and U). Their structures are given in Fig.~\ref{fig:intro:molecules}. The study of the isolated nucleobases in gas phase is only a small piece of a much larger puzzle, but it can provide a unique insight into the behavior of matter and also leads to a general understanding of the relationship between structure and the fate of excited states.

In order to carry out dynamical simulations, the appropriate equations of motion have to be identified and then solved by suitable numerical tools. Therefore, this chapter starts with a short overview of the methods that can be employed to this aim. A number of issues should be kept in mind when trying to connect the results  from dynamical simulations and the available time-resolved experiments; these are discussed in Section~\ref{sec:sim_exp}. Afterwards, we review the different dynamical studies available on the isolated nucleobases (A, G, C, T and finally U), each time starting with a short summary of the experimentally obtained relaxation lifetimes and followed by the different mechanisms that have been suggested to explain these lifetimes. In the conclusions, general trends for purine and pyrimidine bases are compared and the advances of this field are put into perspective. 

\section{Computational Approaches for Nuclear Dynamics}\label{sec:methods}

As pointed out above, understanding the photophysics of DNA nucleobases is not complete without a description of the dynamical processes triggered by UV excitation. In the following section, we review the most popular computational approaches to describe chemical dynamics. The list of methods is not complete, rather the focus is put on those methods which have been employed to calculate time-resolved properties in DNA nucleobases. It is nevertheless not the intention of this section to give a full description of the chosen methods, but only to provide a brief insight into their fundamentals in order to understand put the available simulations in context.

The study of chemical dynamics is the description of nuclear motion. Therefore, this section starts with the fundamental time-dependent Schr\"odinger equation and the Born-Oppenheimer approximation. Afterwards, the most accurate method to describe dynamics, i.e.\ the fully quantum-mechanical approach of wavepacket quantum dynamics (QD) is explained, followed by the multi-configurational time-dependent Hartree (MCTDH) method. Then we will motivate the use of a classical approximation to the nuclear motion, and from there add methodological features which improve the description of excited-state dynamics in full dimensionality. We will discuss the conceptually simple approach of mean-field (MF) dynamics, the widely employed trajectory surface hopping (TSH) scheme and finally the full multiple spawning (FMS) approach.


\subsection{The Schr\"odinger Equation and the Born-Oppenheimer Approximation}

\index{Schr\"odinger equation}

The time-dependent Schr\"odinger equation (TDSE) provides the exact quantum-mechanically and non-relativistically time evolution of a molecule. It reads as
\begin{equation}
  \hat{\mathcal{H}}\left|\Psi^{\mathrm{tot}}\right\rangle
  =
  \I\hbar\frac{\partial}{\partial t}\left|\Psi^{\mathrm{tot}}\right\rangle.
  \label{eq:m:TDSE}
\end{equation}
Here, $|\Psi^{\mathrm{tot}}\rangle$ is the total wavefunction and $\hat{\mathcal{H}}$ is the total Hamiltonian, which contains the kinetic energy of all nuclei and electrons as well as the potential energy arising from the interactions of these particles:
\begin{equation}
  \hat{\mathcal{H}}
  =\hat{\mathcal{T}}^{\mathrm{nuc}}
  +\hat{\mathcal{T}}^{\mathrm{el}}
  +\hat{\mathcal{V}}^{\mathrm{nuc,nuc}}
  +\hat{\mathcal{V}}^{\mathrm{nuc,el}}
  +\hat{\mathcal{V}}^{\mathrm{el,el}}.
  \label{eq:m:MCH}
\end{equation}
Since the motion of the particles is correlated by their mutual interactions, the Schr\"odinger equation
can only be solved exactly for two-particle molecules (i.e.\ the hydrogen atom). Thus,
several approximations are required to describe larger systems, such as molecules.
The most important approximation is probably the Born-Oppenheimer approximation, which allows to separate the nuclear from the electronic motion.

\index{Schr\"odinger equation!electronic}

The electronic wavefunction can be obtained by solving the time-independent electronic Schr\"odinger equation
\begin{equation}
  \hat{\mathcal{H}}^{\mathrm{el}}
  \left|\Psi^{\mathrm{el}}\right\rangle
  =
  E^{\mathrm{el}}
  \left|\Psi^{\mathrm{el}}\right\rangle,
  \label{eq:m:elSE}
\end{equation}
where an electronic Hamiltonian---the so-called clamped-nuclei Hamiltonian---is given by 
\begin{equation}
  \hat{\mathcal{H}}^{\mathrm{el}}
  =\hat{\mathcal{T}}^{\mathrm{el}}
  +\hat{\mathcal{V}}^{\mathrm{nuc,nuc}}
  +\hat{\mathcal{V}}^{\mathrm{nuc,el}}
  +\hat{\mathcal{V}}^{\mathrm{el,el}}.
  \label{eq:m:elH}
\end{equation}
Solving~(\ref{eq:m:elSE}) is the realm of quantum chemistry and a number of powerful electronic structure methods are available in different quantum chemistry codes. For the description of electronic excited states of nucleobases and related properties, the most popular methods have been described in the previous Chapter (Sect.~2.2). In any case, the approximate solutions of the electronic Schr\"odinger equation deliver electronic wavefunctions and electronic energies, which depend parametrically on the nuclear coordinates $\vec{R}$. The functions $E^{\mathrm{el}}(\vec{R})$ are known as potential energy hypersurfaces (PEH).

\index{Born-Oppenheimer approximation}
\index{Non-adiabatic coupling}
\index{Schr\"odinger equation!nuclear}

Within the Born-Oppenheimer approximation the total wavefunction can be written as
\begin{equation}
  \left|
    \Psi^{\mathrm{tot}}
  \right\rangle
  =
  \sum\limits_{\alpha}
  \left|
    \Psi^{\mathrm{el}}_\alpha
  \right\rangle
  \left|
    \Psi^{\mathrm{nuc\phantom{l}}}_\alpha
  \right\rangle.
  \label{eq:m:BOwf}
\end{equation}
By inserting~(\ref{eq:m:BOwf}) and $\hat{\mathcal{H}}=\hat{\mathcal{T}}^{\mathrm{nuc}}+\hat{\mathcal{H}}^{\mathrm{el}}$ in the TDSE~(\ref{eq:m:TDSE}), projecting on $\langle\Psi_\beta^{\mathrm{el}}|$ and using~(\ref{eq:m:elSE})~\cite{Doltsinis2006}, the nuclear Schr\"odinger equation becomes
\begin{equation}
  \left[
    \hat{\mathcal{T}}^{\mathrm{nuc}}
    +E^{\mathrm{el}}_\beta
  \right]
  \left|
    \Psi^{\mathrm{nuc\phantom{l}}}_\beta
  \right\rangle
  +\sum\limits_\alpha
  \hat{\mathcal{T}}^{\mathrm{NAC}}_{\beta\alpha}
  \left|
    \Psi^{\mathrm{nuc\phantom{l}}}_\alpha
  \right\rangle
  =
  \I\frac{\partial}{\partial t}
  \left|
    \Psi^{\mathrm{nuc\phantom{l}}}_\beta
  \right\rangle.
  \label{eq:m:nucSE}
\end{equation}
According to this equation, the nuclei move in the potentials $E^{\mathrm{el}}(\vec{R})$ which are determined by the electronic motion in the field of the nuclei. The non-adiabatic coupling (NAC) operator
\begin{equation}
  \hat{\mathcal{T}}^{\mathrm{NAC}}_{\beta\alpha}
  =
  \sum\limits_a
  \frac{\hbar^2}{m_a}
  \left[
    \left\langle
      \Psi^{\mathrm{el}}_\beta
    \middle|
      \nabla_a^2
    \middle|
      \Psi^{\mathrm{el}}_\alpha
    \right\rangle
    -
    \left\langle
      \Psi^{\mathrm{el}}_\beta
    \middle|
      \nabla_a
    \middle|
      \Psi^{\mathrm{el}}_\alpha
    \right\rangle
    \nabla_a
  \right],
  \label{eq:m:NAC}
\end{equation}
arising from the action of $\hat{\mathcal{T}}^{\mathrm{nuc}}$ on the electronic wavefunctions, describes the coupling between the electronic states in situations where the electronic wavefunction cannot adapt fast enough to the nuclear motion. Within the Born-Oppenheimer approximation, these couplings are completely neglected and the nuclei move according to~(\ref{eq:m:nucSE}) in the potential corresponding to a single electronic state. This approximation is valid in many situations, especially if only the electronic ground state is involved, for example in thermal reactions. In photochemistry and photophysics, chemical reactions involve several electronic states and nuclear dynamics proceed on several PEHs which usually present crossings, where the Born-Oppenheimer approximation breaks down. In these cases, the couplings described in~(\ref{eq:m:NAC}) cannot be neglected anymore.


\subsection{Quantum Dynamics}\label{sssec:methods:dyn:QD}

\index{Quantum dynamics|(}

The most accurate description of nuclear dynamics can be achieved by solving~(\ref{eq:m:nucSE}) numerically. This is known as wavepacket dynamics or full quantum dynamics, here abbreviated as QD.

In order to solve~(\ref{eq:m:nucSE}), the wavefunction $|\Psi^{\mathrm{nuc}}\rangle$  is linearly expanded in terms of time-independent basis functions $|\phi_\mu\rangle$ (usually in a grid, but not necessarily). In one dimension, this expansion reads
\begin{equation}
  \left|
    \Psi^{\mathrm{nuc}}(t)
  \right\rangle
  =
  \sum\limits_\mu^{N^{\mathrm{BF}}}
  c_{\mu}(t)
  \left|
    \phi_\mu
  \right\rangle,
\end{equation}
with $N^{\mathrm{BF}}$ the number of basis functions. In several dimensions, the analogue expression is 
\begin{equation}
  \left|
    \Psi^{\mathrm{nuc}}(R_1,\dots,R_f,t)
  \right\rangle
  =
  \sum\limits_{\mu_1}^{N^{\mathrm{BF}}_1}\dots\sum\limits_{\mu_f}^{N^{\mathrm{BF}}_f}
  c_{\mu_1\dots\mu_f}(t)
  \prod\limits_{\kappa=1}^f
  \left|
    \phi^{(\kappa)}_{\mu_\kappa}(R_\kappa)
  \right\rangle.
  \label{eq:m:qd_wf}
\end{equation}
Here, $f$ is the number of degrees of freedom (DOF), $c_{\mu_1\dots\mu_f}$ are the wavefunction expansion coefficients, and $|\phi^{(\kappa)}_{\mu_\kappa}(R_\kappa)\rangle$ are the time-independent basis functions for degree of freedom $\kappa$. By plugging this wavefunction into~(\ref{eq:m:nucSE}), the equations of motion for the coefficients are obtained and these equations can be solved numerically with standard matrix algebra computations. By solving the TDSE including the couplings described in equation~(\ref{eq:m:NAC}), a rigorous description of photochemical processes can be carried out. QD simulations based on accurate PEHs provide the best description of a dynamical process and can deliver excellent accuracy. However, QD can only be afforded for small systems, since the calculation of the PEHs suffers from a very unfavorable exponential scaling with the number of DOF. Nowadays, QD simulations are able to treat at most 6 DOF with state-of-the-art techniques, which restricts these calculations to at most 4-atomic molecules. For larger systems, such as DNA nucleobases, only simulations in reduced dimensionality would be feasible. The problem of choosing the appropriate DOF is then not trivial. Particularly, for DNA nucleobases---where a large number of degrees of freedom are involved along the many deactivation pathways (see Chapter~4)---it is very challenging to choose a small subset of reaction coordinates which describe all deactivation pathways reasonably well. This problem motivates the use of more approximate approaches, as the ones presented below.

\index{Quantum dynamics|)}


\subsection{Multi-Configurational Time-Dependent Hartree}\label{sssec:methods:dyn:MCTDH}

\index{Multi-configurational time-dependent Hartree|(}

The multi-configurational time-dependent Hartree (MCTDH) method~\cite{Meyer2009} is an approximation to full QD simulations that allows to considerably extend the applicability of QD to systems with more than 4 atoms. MCTDH is still based on the exact nuclear Schr\"odinger equation, but approximations are introduced in the definition of the wavefunction. The general wavefunction is here given by 
\begin{equation}
  \left|
    \Psi^{\mathrm{nuc}}(R_1,\dots,R_f,t)
  \right\rangle
  =
  \sum\limits_{\mu_1}^{n^{\mathrm{BF}}_1}\dots\sum\limits_{\mu_f}^{n^{\mathrm{BF}}_f}
  c_{\mu_1\dots\mu_f}(t)
  \prod\limits_{\kappa=1}^f
  \left|
    \phi^{(\kappa)}_{\mu_\kappa}(R_\kappa,t)
  \right\rangle.
  \label{eq:m:mctdh_wf}
\end{equation}
The main difference to~(\ref{eq:m:qd_wf}) is the time-dependence of the basis functions. Hence, the equations of motion in MCTDH have to be solved not only for the wavefunction coefficients $c_{\mu_1\dots\mu_f}(t)$, but also for the basis functions $|\phi^{(\kappa)}_{\mu_\kappa}(R_\kappa,t)\rangle$. This means, however, that the basis set expansion can be kept much smaller than in full QD calculations ($n^{\mathrm{BF}}_i \ll N^{\mathrm{BF}}_i$), since in MCTDH the basis functions can adapt during the dynamics. In some cases~\cite{Meyer2009}, only a handful of basis functions are necessary to obtain qualitatively correct results. Obviously, by increasing the number of basis functions, MCTDH approaches the accuracy of full QD calculations, however, in this limit MCTDH becomes as expensive as full QD.

The formal scaling of MCTDH remains exponential with respect to the number of DOF. However, through its definition of the basis functions, the method can be applied to systems involving between 20 and 50 DOF, see e.g.~\cite{Meyer2009}. Furthermore, related techniques as Gaussian-based MCTDH~\cite{Burghardt1999JCP}, Multilayer MCTDH~\cite{Wang2003JCP} and variational Multi-Configurational Gaussians (vMCG)~\cite{Lasorne2006CPL} are very promising to treat much larger systems quantum-mechanically, exemplified by the simulation of the anthracene cation with 66 DOF~\cite{Meng2013JCP}. 

\index{Multi-configurational time-dependent Hartree|)}


\subsection{Molecular Dynamics}\label{sssec:methods:dyn:MD}

\index{Molecular dynamics|(}
\index{Molecular dynamics!ab initio}

While the above methods fully preserve the quantum-mechanical nature of the nuclear dynamics---describing all quantum effects like interference, coherence, and tunneling---they suffer from an unfavorable scaling with system size, i.e.\ the number of DOF. An alternative strategy to describe dynamics is to impose the classical approximation for the motion of the nuclei. Instead of being described by wavepackets moving according to the nuclear Schr\"odinger equation~(\ref{eq:m:nucSE}), the nuclei are treated as classical, point-like particles, which follow Newton's equation of motion
\begin{equation}
  m_a\ddot{\vec{R}}_a=-\nabla_aE^{\mathrm{el}}_\alpha.
  \label{eq:m:newton}
\end{equation}
The force acting on the nuclei is the gradient of the PEH of a single electronic state $\alpha$, in the frame of the Born-Oppenheimer approximation. Such a calculation is usually known as a molecular dynamics (MD) simulation. In the case that the energies $E^{\mathrm{el}}_\alpha$ and driving forces $-\nabla_aE^{\mathrm{el}}_\alpha$ are calculated by means of quantum-chemical electronic structure methods, the simulations are oftentimes called ab initio MD (AIMD), or semi-classical MD (since the electrons are described quantum-mechanically and the nuclei classically).

\index{Molecular dynamics!on-the-fly}

The \emph{global} computation of the potential function $E^{\mathrm{el}}_\alpha$ (prior to the dynamics simulation) still scales exponentially with the number of DOF. Thus, MD simulations are usually performed with ``on-the-fly'' calculations of the potential energy and nuclear forces at each time step of the simulation. In this way, the simulation cost does not explicitly depend on the number of DOF anymore and thus there is no need to restrict the calculation to a certain subset of reaction coordinates. 

\subsection{Ehrenfest Dynamics}\label{sssec:methods:dyn:MF}

\index{Ehrenfest dynamics|(}

Without any further extension, semi-classical MD cannot describe excited-state dynamics because the classical nuclei are tied to one single Born-Oppenheimer PEH at all times. In photophysics and photochemistry, several electronic states are close in energy and interact via the NACs given in ~(\ref{eq:m:NAC}) during internal conversion (IC) close to conical intersections (CoIn), or during intersystem crossing (ISC) via spin-orbit coupling (SOC). At these interstate crossing points on the PEHs, population can be transferred from one state to the other, which is obviously not possible within MD simulations as explained above.

The Ehrenfest dynamics method, also called mean-field (MF) dynamics, is an extension of classical MD calculations that includes excited-state PEHs. Here, the electronic wavefunction is a linear combination of several electronic states
\begin{equation}
  \left|
    \Psi^{\mathrm{el}}
  \right\rangle
  =\sum\limits_{\alpha}
  c_\alpha\left|\Psi_\alpha\right\rangle.
  \label{eq:m:md_elwf}
\end{equation}
The wavefunction coefficients are propagated along with the trajectory according to the energies and the NACs of the electronic states. The potential energy for the MD simulation is substituted with the energy expectation value of the electronic wavefunction
\begin{equation}
  E^{\mathrm{eff}}
  =
  \left\langle
    \Psi^{\mathrm{el}}
  \middle|
    \hat{\mathcal{H}}^{\mathrm{el}}
  \middle|
    \Psi^{\mathrm{el}}
  \right\rangle
  =
  \sum\limits_\alpha
  |c_\alpha|^2E_\alpha^{\mathrm{el}}.
\end{equation}
Thus, the nuclei are moving on an effective potential, which is the average of all adiabatic states of the same multiplicity, weighted by their state population, giving the method the name mean-field dynamics.

The mean-field approach suffers from the problem that the trajectory may follow a nonphysical mixed state after passing a non-adiabatic coupling region. A physical correct description would describe a splitting of the population into different reaction channels. 

\index{Ehrenfest dynamics|)}


\subsection{Trajectory Surface Hopping}\label{sssec:methods:dyn:TSH}

\index{Trajectory surface hopping|(}

As mentioned above, the main problem of the Ehrenfest dynamics is that it cannot describe a wavepacket splitting onto several PEHs. In order to get rid of this shortcoming, the trajectory surface hopping (TSH) scheme~\cite{Tully1971JCP} was devised. Here, the mean-field trajectory is replaced by an ensemble of many trajectories, each following the classical equations of motion. Non-adiabatic effects are described by allowing the trajectories to switch stochastically between the PEHs, based on the strength of the NACs. Using a sufficiently large ensemble of trajectories the splitting of a wavepacket due to non-adiabatic interactions can be approximated.

\index{Trajectory surface hopping!fewest-switches criterion}

Similarly to~(\ref{eq:m:md_elwf}), in the TSH approach the electronic wavefunction is expanded in the basis of the electronic states. The absolute square of the complex coefficients $|c_\alpha|^2$ can be interpreted as the probability of finding the trajectory in state $\alpha$. Thus, from the coefficients one can derive an expression for the instantaneous probability $P_{\beta\rightarrow}$ of leaving the currently occupied classical state $\beta$. According to Tully's fewest switches criterion~\cite{Tully1990JCP}, this probability is given by
\begin{equation}
  P_{\beta\rightarrow}
  =-\frac{2\Delta t}{|c_\beta|^2}
  \Re
  \left(
    c_\beta^*
    \frac{\partial}{\partial t}
    c_\beta
  \right).
  \label{eq:m:sh_prob1}
\end{equation}
The fewest switches criterion states that the surface hopping probabilities should minimize the number of hops while maintaining consistency between the population $|c_\alpha|^2$ and the fraction of trajectories assigned to the state $\alpha$.

The time-derivatives of the coefficients in equation~(\ref{eq:m:sh_prob1}) are directly obtained from the equation of motion of the coefficients
\begin{equation}
  \frac{\partial}{\partial t}
  c_\beta
  =-\sum\limits_\alpha
  \left[
    \frac{i}{\hbar}H_{\beta\alpha}
    +\vec{v}
    \cdot
    T_{\beta\alpha}^{(1)}
  \right]
  c_\alpha,
  \label{eq:m:sh_eom}
\end{equation}
where $\vec{v}$ is the nuclear velocity vector and where 
\begin{equation}
  \vec{v}\cdot
  T_{\beta\alpha}^{(1)}
  =
  \left\langle
    \psi_\beta
    \middle|
    \frac{\partial}{\partial t}
    \middle|
    \psi_\alpha
  \right\rangle.
\end{equation}
The quantities $H_{\beta\alpha}$ and $T_{\beta\alpha}^{(1)}$ are calculated on-the-fly along with the gradient of the populated state. Equation~(\ref{eq:m:sh_eom}) can be integrated by standard Runge-Kutta methods or unitary propagator methods, using small time steps $\Delta t$. Finally, by inserting~(\ref{eq:m:sh_eom}) into~(\ref{eq:m:sh_prob1}), the probability $P_{\beta\rightarrow}$ can be written as a sum of $P_{\beta\rightarrow\alpha}$, the probability of switching from state $\beta$ to state $\alpha$.

\index{Molecular dynamics!Car-Parinello dynamics}

Besides Tully's fewest switches criterion, there exist additional approaches to calculate the TSH probabilities, like coherent switching with decay of mixing~\cite{Zhu2004JCP} or fewest switches with time-uncertainty~\cite{Jasper2002JCP}; also combination of Car-Parinello molecular dynamics (CPMD) with TSH (TSH-CP)~\cite{Doltsinis2002PRL} and the combination of TSH with mean-field dynamics~\cite{Prezhdo1997JCP} have been reported.

\index{Trajectory surface hopping!\textsc{Sharc}}
\index{Spin-orbit coupling}
\index{Intersystem crossing}

The TSH scheme described above has been only employed to study the photochemical deactivation pathways involving  PEHs of the same multiplicity, i.e. via internal conversion. Recently~\cite{Richter2011JCTC}, the TSH scheme has been extended to treat ISC mediated by SOC. The SOC matrix elements appear as off-diagonal elements in $H_{\beta\alpha}$ in~(\ref{eq:m:sh_eom}), which couple states of different multiplicity. Without any further changes, Eq.~(\ref{eq:m:sh_eom}) could be propagated including SOC in $H_{\beta\alpha}$ (this is sometimes called ``spin-diabatic approach''~\cite{Granucci2012JCP}), but this scheme is not rotationally invariant and neglects the effect of the SOC on the PEHs (zero-field splitting). Additionally, the TSH scheme is based on the assumption that the couplings between the electronic states are localized (as the NACs around a CoIn are), while SOCs are clearly not. Within the Surface Hopping including Arbitrary Couplings (\textsc{Sharc}) methodology~\cite{Richter2011JCTC}, the Hamiltonian is diagonalized, yielding fully adiabatic, spin-mixed electronic states. In this spin-mixed basis, the non-local SOCs are transformed into localized non-adiabatic couplings, which allows to use the TSH method in the intended way. The diagonalization can be written in terms of a unitary transformation between the electronic states,
\begin{equation}
  \vec{U}^\dagger\vec{H}\vec{U}
  =\vec{H}^{\mathrm{diag}}.
  \label{eq:m:sharc_h}
\end{equation}
Here, $\vec{H}$ is the Hamiltonian matrix represented in the basis of the eigenfunctions of the molecular Coulomb Hamiltonian (given in~(\ref{eq:m:MCH})), $\vec{H}^{\mathrm{diag}}$ is the same matrix in diagonalized form, and $\vec{U}$ is the unitary transformation matrix. In order to consistently include the non-adiabatic couplings, they have to be transformed as well. This leads to a new equation of motion for the coefficients in the diagonal basis:
\begin{equation}
  \frac{\partial}{\partial t}
  c_\beta^{\mathrm{diag}}
  =-\sum\limits_\alpha
  \left[
    \frac{i}{\hbar}
    \left(
      \vec{U}^\dagger\vec{H}\vec{U}
    \right)_{\beta\alpha}
    +\vec{v}
    \cdot
    \left(
      \vec{U}^\dagger\vec{T}^{(1)}\vec{U}
    \right)_{\beta\alpha}
    -\left(
    \vec{U}^\dagger\frac{\partial\vec{U}}{\partial t}
    \right)
  \right]
  c_\alpha^{\mathrm{diag}}.
  \label{eq:m:sharc_eom}
\end{equation}
From this equation, the surface hopping probabilities can be calculated, in a similar fashion as in the original TSH approach. Since in \textsc{Sharc} the nuclei follow spin-mixed PEHs, the spin-mixed gradients have to be calculated from the gradients of the unmixed states and the eigenvectors given by the matrix $\vec{U}$. In the current approach, this neglects the derivatives of the SOC elements with respect to the nuclear coordinates, but this is a reasonable good approximation because the SOC operator is of short-ranged nature~\cite{Hess1996CPL}. Since in \textsc{Sharc} gradients for several electronic states have to be calculated in order to evaluate the mixed gradient, this approach is slightly more expensive than regular TSH. The price is, however, the money worth, since the inclusion of ISC processes can be as relevant as those processes mediated by internal conversion, even in systems with light atoms such as nucleobases.

TSH, in any of many different flavours, is the most widely used method employed to perform excited-state dynamics of DNA nucleobases. Compared to QD methods, where the number of electronic structure calculations scales exponentially with the number of DOF, in TSH methods the energies, forces and couplings are evaluated on-the-fly, and thus the number of electronic structure only depends on the number of trajectories and the desired number of time steps. This gives TSH simulations the enormous advantage of being able to include all molecular degrees of freedom in the simulation, even for relatively large molecules. In fact, the cost of TSH simulations is basically only dependent on the cost of the on-the-fly electronic structure calculations as the cost of integrating~(\ref{eq:m:newton}) and~(\ref{eq:m:sh_eom}) is almost negligible. An additional advantage of TSH methods is that---since the trajectories are all independent of each other---they can be executed in parallel, further increasing the efficiency of the approach.

Despite its attractiveness, one has to keep in mind that due to its semi-classical nature, TSH fails to properly describe a number of quantum effects. First, the method cannot account for tunneling of the intrinsically classical nuclei; a description of tunneling for selected DOF is still possible~\cite{Hammes-schiffer1994JCP}. Another shortcoming of TSH methods is that quantum coherences between the electronic states are often not described correctly. One possibility is to add the so-called decoherence corrections~\cite{Schwartz1996JCP,Granucci2007JCP}. For more details on the TSH method, see the excellent review in~\cite{Barbatti2011WCMS}.
 
\index{Trajectory surface hopping|)}
\index{Molecular dynamics|)}
 

\subsection{Full Multiple Spawning}\label{sssec:methods:dyn:FMS}

\index{Full multiple spawning|(}

The Full Multiple Spawning (FMS) methodology~\cite{Martinez1996JPC,Martinez1997JPCA,Ben-nun2002} could be considered a step from TSH towards full quantum-mechanical calculations. However, as it requires only local knowledge about the PEHs, it is still suited for on-the-fly calculations. FMS coupled to on-the-fly ab initio quantum chemistry is usually termed ab initio multiple spawning (AIMS). In a nutshell, in FMS the nuclear wavefunction is expanded in a basis of frozen Gaussians, whose centers follow classical trajectories. Non-adiabatic effects are described by population transfer between basis functions assigned to different electronic states. To minimize the size of the Gaussian basis set while accurately describing a wavepacket, new Gaussians are spawned whenever necessary.

Within FMS, the total wavefunction is expanded as in the Born-Oppenheimer ansatz (see~(\ref{eq:m:BOwf})).
\begin{equation}
  \left|
    \Psi^{\mathrm{tot}}(\vec{R},\vec{r},t)
  \right\rangle
  =
  \sum\limits_{\alpha=1}
  \left|
    \Psi^{\mathrm{el}}_\alpha(\vec{r};\vec{R})
  \right\rangle
  \left|
    \Psi^{\mathrm{nuc\phantom{l}}}_\alpha(\vec{R},t)
  \right\rangle
\end{equation}
Each nuclear wavefunction is represented in a time-dependent basis set composed of Gaussian basis functions $G^\alpha_\mu$:
\begin{equation}
  \left|
    \Psi^{\mathrm{nuc\phantom{l}}}_\alpha(\vec{R},t)
  \right\rangle
  =
  \sum\limits_\mu^{N_\alpha(t)}
  c^\alpha_\mu(t)
  G^\alpha_\mu\left[
    \vec{R}; \bar{\vec{R}}^\alpha_\mu(t),\bar{\vec{P}}^\alpha_\mu(t),\gamma^\alpha_\mu(t)
  \right],
\end{equation}
where $c^\alpha_\mu(t)$ are the complex coefficients, and there are $N_\alpha(t)$ basis functions for each electronic state $\alpha$ at time $t$. The Gaussian basis functions are specified by an average position $\bar{\vec{R}}(t)$, an average momentum $\bar{\vec{P}}(t)$, a phase factor $\gamma^\alpha_\mu(t)$ and by a set of time-independent width parameters for each degree of freedom (i.e.\ the basis functions are frozen Gaussians).

In FMS, the parameters $\bar{\vec{R}}(t)$ and $\bar{\vec{P}}(t)$ are propagated according to Newton's equation of motion~(\ref{eq:m:newton}), very similar to the surface hopping method. In a region of strong non-adiabatic coupling new basis functions are spawned on the coupled surface. In this fashion, the splitting of a wavepacket and the involved population transfer in such a region can be modelled accurately.

Similar to TSH methods, the total cost of FMS calculations is primarily determined by the cost of the on-the-fly electronic structure calculations. However, since new basis functions are spawned every time a region of non-adiabatic effects is traversed, FMS is more expensive than TSH. On the positive side, FMS allows for the correct quantum-mechanical description of coherences between the different parts of the nuclear wavepacket and combined with high-level correlated electronic structure methods~\cite{Coe2007JPCA} it may provide quantitatively correct results. According to Hack et al.~\cite{Hack2001JCP}, FMS delivers results which are as good or better than TSH calculations.

\index{Full multiple spawning|)}


\section{Connection of the Dynamics Simulations to Experiment}\label{sec:sim_exp}

The above explained theoretical methods aim at describing the real dynamical processes detected in an experiment. The complexity of these processes, however, affects what one can learn from both theory and experiments. On the one hand, simulations require approximations in order to be capable to treat the considered systems and these approximations need to be validated by experiments. On the other hand, the experimental signals are extremely difficult to interpret without the help of computational predictions. Therefore, theory and experiment need to act in concert.

It might seem obvious that it is desirable to simulate all the processes induced in an experiment and yield directly comparable results. However, this simple wish is not easy to fulfill, since in most cases the measured observables and the ones calculated are not the same. Typically, only the evolution of the excited-state population is calculated. However, an experiment consists of excitation (pumping), evolution (excited-state dynamics), probing, side effects induced by the probing procedure like fragmentation, and finally detection of a signal. The interpretation of this signal by means of theoretical simulations requires to know which of these phenomena are really simulated and how. To help this task, it is useful first to understand the different experimental setups and then make a connection with the particular computational approaches.

The different experimental approaches can be categorized on different levels. One can distinguish between gas phase and liquid phase experiments but also between ionization, transient absorption or fluorescence techniques. A detailed description can be found in Chapters 2, 3, 8, 10, 12, 15, 16 and also in the reviews~\cite{Staniforth2013PRSA, Kleinermanns2013IRPC, Middleton2009ARPC, Markovitsi2007PPS, Shukla2007JBSD, Saigusa2006JPPC, Crespo-Hernandez2004CR, Fischer2003CSR}. Here, we shall focus only on the elements needed to do a sensible connection between the experiments and gas-phase simulations. The common essence of all possible spectroscopic experiments is that the nucleobase is first excited by UV light, stemming typically from an ultrashort laser pulse. In this way, excited-state dynamics is initiated, which is probed after a variable delay. The different applicable experimental techniques will condition the information obtained about the excited-state populations. 

In the following, we summarize several points that should be kept in mind when simulating dynamical processes. 

\subsubsection*{The Timescale}

Using the methods described in Sect.~\ref{sec:methods}, only ultra-short dynamics (below a few ps) can be computed. This limitation is given by the high computational cost of the electronic structure calculations at each time step. Additionally, over the course of many time steps errors accumulate, making long simulations particularly error-prone.

\subsubsection*{The System Size}

The size of the system considered in the simulation plays an important role. A calculation of an isolated nucleobase strictly corresponds to a gas-phase experiment but may also be used as an approximation for liquid-phase experiments. Improvements in order to account for solvent effects are possible, see e.g.\ Chap.~9, but usually involve additional approximations and computational cost.

\subsubsection*{The Excitation Process}

In the majority of calculations, the UV excitation is not directly simulated but approached in an ad-hoc fashion, i.e.\ by starting the simulation with population in a carefully selected distribution of excited states. This implies the use of an infinitely short pump pulse (a $\delta$-pulse) while in reality the experimental pump pulse has a finite duration. This approximation is usually of little importance in comparison with other approximations and thus the experimental pump process is reasonably well described.
 
\subsubsection*{The Quality of the Potential Energy Hypersurfaces}

The excited-state PEHs, on which the dynamics is simulated, are calculated with electronic structure methods that have an associated limited accuracy. A wide variety of methods exist, ranging from semi-empirical to high-level multi-reference (MR) approaches. Most of the computational studies on nucleic acids use one of the following methods: TD-DFT (time-dependent density functional theory), DFTB (density functional based tight binding), CASSCF (complete active space self-consistent field), CASPT2 (complete active space perturbation theory of second order), MRCI (multi-reference configuration interaction) and CI methods based on semi-empirical Hamiltonians like OM2 or AM1. These methods differ in their ability to properly describe excited-state PEHs, with more accurate methods usually being more expensive, see e.g.~\cite{Szabo1996,Helgaker2000,Levine2001,Cramer2004,Jensen2007}. Even highly sophisticated methods involve approximations which need to be validated by experimental results. Commonly, this validation involves a comparison of the calculated excitation energies at the ground state geometry with the experimental absorption spectrum. The experimental spectrum is often broad with overlapping bands, making the assignment of the computed excited-state energies far from straightforward. Additionally, one has to keep in mind that a method giving good results at the equilibrium geometry still may perform badly at other non-equilibrium geometries. In most of the cases, a further validation using experimental data is not possible apart from a comparison of the actual dynamics results. Another limiting factor when choosing a method is that some properties (e.g.\ gradients of the potential energy) needed in certain types of dynamics simulations may be simply not available for various sophisticated electronic structure methods. As a consequence, the choice of a method is a compromise between accuracy, method availability and computational cost.

\subsubsection*{The Representation}

The choice of the representation, in which state populations are computed, can be of great importance. The representation which arises naturally from the Born-Oppenheimer approximation and which is commonly used in electronic structure calculations, is termed ``adiabatic'' representation. Here, electronic states are ordered strictly according to energy and population transfer between these states is introduced by the so-called NACs in the kinetic part of the Hamiltonian (here abbreviated as k-NAC). The adiabatic singlet states are usually referred to as $S_0$, $S_1$ and so on, with similar labels for other multiplicities. Adiabatic states do not preserve the wavefunction character. As a consequence, properties like the transition dipole moment may change drastically along a coordinate involving a fast change of wavefunction character.

In contrast to the adiabatic representation, states preserve their wavefunction character in the so-called ``diabatic'' representations (note that an infinite number of diabatic representations exists). In opposition to adiabatic states, PEHs of diabatic states can cross and molecular properties are usually smooth functions of the internal coordinates. The nomenclature of these states is often derived from the dominant excitation with respect to the ground state, e.g.\ $\pi\pi^*$ or $n\pi^*$. By changing from the adiabatic to the diabatic representation, the NACs in the kinetic part of the Hamiltonian are transformed into potential couplings (here p-NAC).

Quantum mechanically, both representations are strictly equivalent and correct as long as all couplings between all relevant states are considered. However, the state populations expressed in different representations can differ significantly. As a consequence, lifetimes obtained from fitting the excited-state populations are also dependent on the chosen representation. Since in the diabatic representation molecular properties change smoothly along a given coordinate, it is the most suitable representation for comparison with the experiment, here we term this representation ``spectroscopic''. As mentioned above, electronic structure calculations usually yield energies in the adiabatic representation and a transformation to the diabatic or spectroscopic picture is by no means trivial.

\index{Non-adiabatic coupling}
\index{Spin-orbit coupling}

The problem of choosing the correct representation for dynamics simulations becomes more complicated if SOCs are involved. Then, states of different multiplicity can mix and we can distinguish between representations where spin-orbit couplings are introduced either as potential couplings (p-SOC) or as kinetic couplings (k-SOC). The corresponding potentials are sometimes termed ``spin-free'' or ``spin-diabatic'' in the case of p-SOC and ``spin-mixed'' or ``spin-adiabatic'' in the case of k-SOC~\cite{Richter2011JCTC,Granucci2012JCP}. The standard electronic structure programs yield p-SOC. Thus, the outcome of ab initio codes is usually adiabatic with respect to the NAC (k-NAC) but diabatic with respect to the SOC (p-SOC). To avoid the confusion between the terms ''diabatic`` and ``adiabatic'', the following terminology~\cite{Mai2013CPC} might be more convenient: In the ``spectroscopic'' (superscript spec) representation, the wavefunction character is preserved and the picture is closest to spectroscopic results. The ``molecular Couloumb Hamiltonian'' (MCH) representation is the standard output of quantum chemistry programs, where NACs are calculated as kinetic couplings and SOCs are off-diagonal terms of the potential when using a matrix form. Finally, the ``diagonal'' (diag) representation means that a fully diagonal potential matrix is obtained. Here, states of different multiplicity mix and terms like $S$ (singlet) and $T$ (triplet) might not be meaningful anymore. Instead, only the energetic ordering is unambiguous. For a distinction of the three representations, we write the corresponding total Hamiltonians in a simplified two-potential system:
 \begin{equation}
   \hat{\mathcal{H}}^{\mathrm{spec}} =
   \left( \begin{array}{cc}
      T^{\mathrm{spec}}_1 & 0 \\
      0                   & T^{\mathrm{spec}}_2
   \end{array} \right) +
  \left( \begin{array}{cc}
      V^{\mathrm{spec}}_1             & \mathrm{p-NAC} + \mathrm{p-SOC} \\
      \mathrm{p-NAC} + \mathrm{p-SOC} & V^{\mathrm{spec}}_2
   \end{array} \right)
 \end{equation}
 \begin{equation}
   \hat{\mathcal{H}}^{\mathrm{MCH}} =
   \left( \begin{array}{cc}
      T^{\mathrm{MCH}}_1 & \mathrm{k-NAC} \\
      \mathrm{k-NAC}      & T^{\mathrm{MCH}}_2
   \end{array} \right) +
  \left( \begin{array}{cc}
      V^{\mathrm{MCH}}_1 & \mathrm{p-SOC} \\
      \mathrm{p-SOC}      & V^{\mathrm{MCH}}_2
   \end{array} \right)
 \end{equation}
 \begin{equation}
   \hat{\mathcal{H}}^{\mathrm{diag}} =
   \left( \begin{array}{cc}
      T^{\mathrm{diag}}_1             & \mathrm{k-NAC} + \mathrm{k-SOC} \\
      \mathrm{k-NAC} + \mathrm{k-SOC} & T^{\mathrm{diag}}_2
   \end{array} \right) +
  \left( \begin{array}{cc}
      V^{\mathrm{diag}}_1 & 0 \\
      0                  & V^{\mathrm{diag}}_2
   \end{array} \right)
 \end{equation}
We note here that in semiclassical simulations the representations are not fully equivalent~\cite{Tully1990JCP,Mai2013CPC} and thus special care is required.

\begin{figure}
  \includegraphics[width=\textwidth]{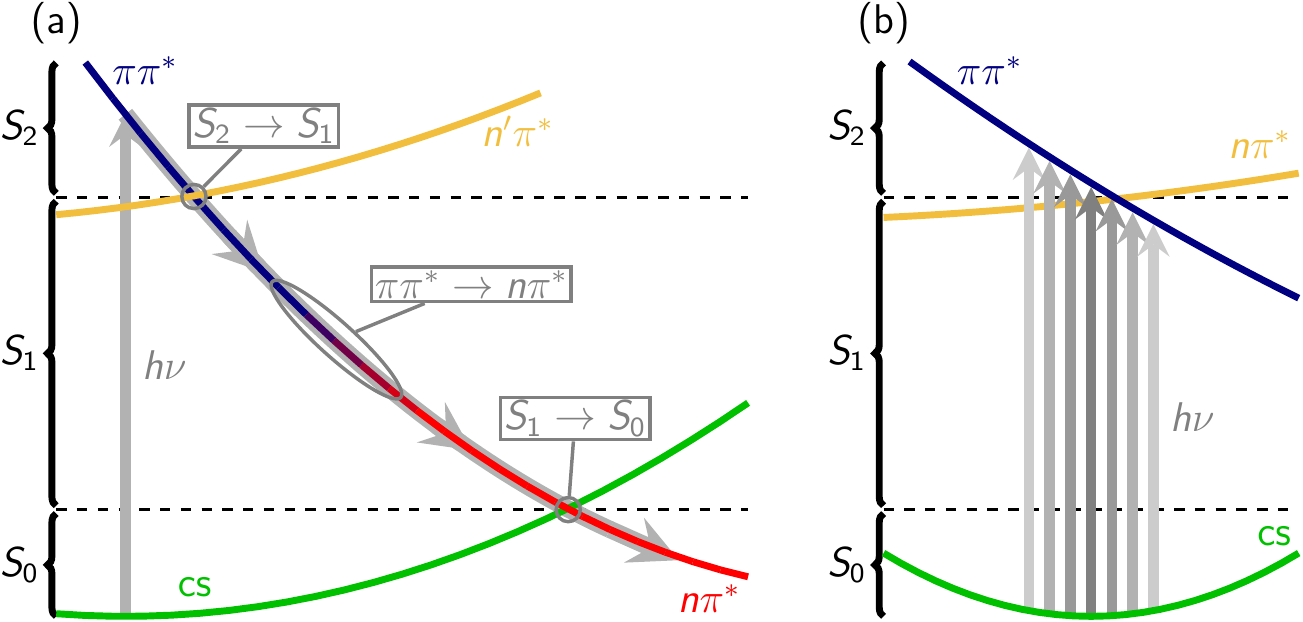}
  \caption{Comparing representations: \textbf{a} A relaxation pathway after photoexcitation (grey arrows) is shown in a simple model. The underlying potential energy curves are labeled according to their predominant state character ($\pi\pi^*$ \emph{blue}, $n'\pi^*$ \emph{yellow}, $n\pi^*$ \emph{red}, $cs$ (closed shell) \emph{green}; spectroscopic representation) or their energetic ordering ($S_2$, $S_1$, $S_0$, note the \emph{dashed} separation lines; MCH representation). Crossings are indicated with circles and labeled according to the involved states. \textbf{b} Photoexcitation involves several geometries and depending on the representation, different states can be excited, if a CoIn is located in the Franck-Condon region.}
  \label{fig:22:representations}
\end{figure}

In the following, we illustrate the general discussion on the representations with two examples. First, we focus on the difference between the spectroscopic and the MCH representation. In Fig.~\ref{fig:22:representations}a, an exemplary excitation-relaxation pathway is schematically depicted. Population is vertically excited from the ground state (closed-shell, cs) minimum and then evolves on different potentials. According to the MCH picture (which is typically employed in semiclassical calculations), the system follows an $S_2 \rightarrow S_1 \rightarrow S_0$ path. Analyzing dynamics results in this representation would result in two distinct time constants, with a small constant associated to the $S_2\rightarrow S_1$ process and larger one to $S_1\rightarrow S_0$. When analyzing the dynamics instead in terms of spectroscopic states the same path would be identified as $\pi\pi^* \rightarrow n\pi^*$ (indicated by the color gradient from blue to red). Accordingly, only one decay time of intermediate duration would be obtained. An experiment where $\pi\pi^*$ and $n\pi^*$ give rise to signals of different strengths would similarly only measure a single time constant. However, a direct comparison of the MCH populations with the experimental transient is not possible. Nevertheless, time constants derived from MCH populations can seemingly agree with experimental time constants due to error compensation. Moreover, the overlap of several competing processes can give rise to effective time constants which might coincide with time constants from MCH populations.
In any case, the most straightforward comparison of simulation and experiment could be obtained by explicitly including the pump and probe processes in the simulation. 

In the second example, one of the reasons for the above-mentioned occurrence of several simultaneous processes is illustrated (see Fig.~\ref{fig:22:representations}b). The Franck-Condon (FC) region, from where excitation takes place, comprises several different geometries around the $S_0$ minimum. A simplification is often used and---especially in static simulations---only the ground state equilibrium geometry is considered (center arrow in the Figure). At this specific geometry, the $S_2$ corresponds to the $\pi\pi^*$ in the present example. Here, the $\pi\pi^*$ is the bright state, where population is transferred from the ground state, while the dark $n\pi^*$ state corresponds to the $S_1$. If a CoIn between the two states is situated in the Franck-Condon region, other geometries exist, where the correspondence between $S_1, S_2$ and $\pi\pi^*, n\pi^*$ is reversed. In other words, the bright state, despite being always the $\pi\pi^*$ state, corresponds to the $S_2$ for some geometries and to the $S_1$ for others. Therefore a correct simulation must involve both states. To complicate things further, the oscillator strength of the $n\pi^*$ state may be small but non-zero. This again imposes that many starting geometries and states might be necessary to fully understand the dynamics of the system. Because all these initial conditions may lead to different processes and outcomes, the dynamical picture might be rather complicated.

\subsubsection*{The Probe Process}

Finally, also the probe process is often not simulated, which can have several reasons. The focus of interest is the dynamics in the excited states, which can be observed only indirectly in an experiment by using a probe. In contrast, this information can be directly accessed in a calculation. Therefore, one might think that the simulation of the probe process is superfluous. However, when one tries to compare theory and experiment, computational state lifetimes are related to signal decay times. On the one hand, these signal decay times are usually ``blind'' for some molecular processes due to a specific experimental setup. Additionally, the probe process used for obtaining the decay times will have some intrinsic limitations, e.g.\ employing a limited energetic window. Therefore, a signal decay time can stem from different state lifetimes and setup-specific effects like running out of the probe window. On the other hand, the approximations made in the simulations may lead to errors in the computed state lifetimes. Hence, great care has to be taken when comparing experimental with theoretical results and it is desirable to simulate the probe process in order to arrive at really analoguous outcomes.

\begin{figure}\label{fig:22:exp}
  \includegraphics[width=\textwidth]{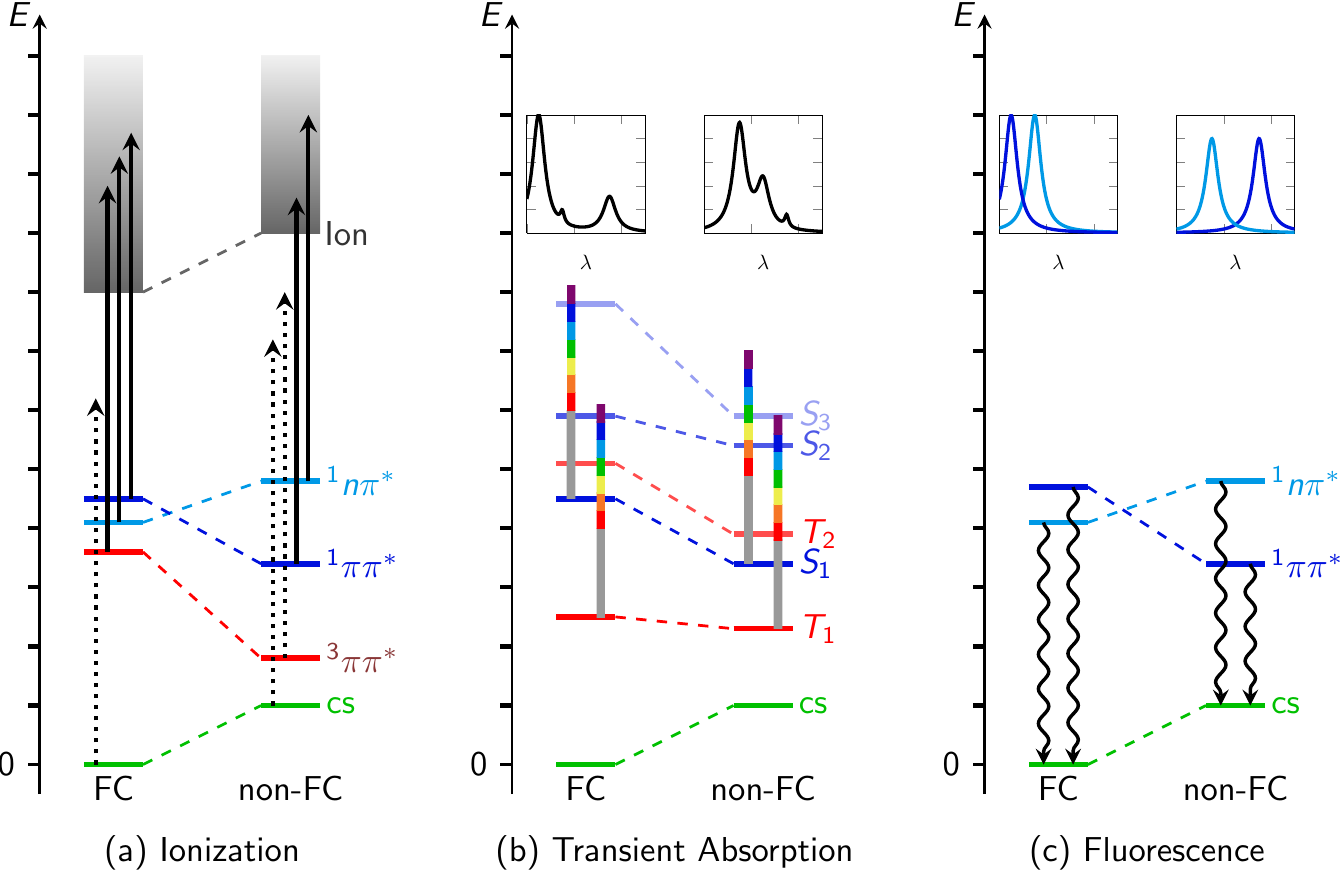}
  \caption{Basics of different experimental probing techniques: 
  \textbf{a} Ionization. The probe pulse has to carry enough energy to ionize the molecule, which is the case for the excited states $^1n\pi^*$, $^1\pi\pi^*$ and $^3\pi\pi^*$ (ionization indicated by \emph{solid} lines), but not for the closed-shell (cs) ground state (ionization not possible, \emph{dotted} line). At geometries far from the FC region, even some excited states (here the $^3\pi\pi^*$) might be too low in energy to be ionized. \textbf{b} Transient absorption. The absorption of light by populated excited states leads to bands in the absorption spectrum. At another geometry, the energies of the excited states will have shifted, giving rise to a different spectrum. \textbf{c} After photoexcitation, the fluorescence is detected in a time-resolved way. During the dynamics, the fluorescence spectrum changes according to the state energies.}
\end{figure}

In what follows, some key concepts of different probing schemes are illustrated and connections to possible simulations of such experiments are made. Here, we limit ourselves to setups using ionization, transient-absorption, or fluorescence (Fig.~\ref{fig:22:exp}). 

Ionization will be discussed first (Fig.~\ref{fig:22:exp}a). After the pump pulse, the excited states get populated. Excited-state population can be detected by ionizing with a probe pulse carrying sufficient energy. Since the pulse does not carry enough energy to ionize the ground state (indicated by the dotted arrow), the signal will be proportional to the population in the excited states only. By varying the delay between pump and probe, the evolution of the excited-state population can be tracked. However, depending on the geometry, additional states---here for example a $^3\pi\pi^*$ triplet---might be too low in energy to be ionized by the probe laser. Thus, a distinction between molecules relaxing to the ground state and molecules being trapped in the low-lying state is not possible anymore. The time constants extracted from the transient are hence an effective time constant including both processes. An additional issue to keep in mind is that the ionization probability depends not only on the energetic separation of the neutral and ionic states, but also depends on the wavefunction character. However, a clear distinction between ionization from a triplet or from a singlet state is extremely difficult. Last but not least, multiphoton processes are often employed in the experimental probing, so that the ionization probability is additionally influenced by the presence of intermediate states, which further complicates the connection between the signal obtained and the actual photorelaxation in the neutral states. While first attempts of simulating ionization yields have been made, see e.g.~\cite{Hudock2007JPCA}, a full simulation of multiphoton ionization is very demanding~\cite{Kotur2012PRL,Spanner2012PRA,Spanner2013CP,Spanner2009PRA} and cannot be carried out on-the-fly during the excited-state dynamics computations.

As a second approach, particularly suited for liquid-phase experiments, transient absorption detection will be briefly discussed (Fig.~\ref{fig:22:exp}b). Here, the excited-state dynamics is ideally probed within a very wide spectral range (indicated by the vertical spectrally colored bars in the Figure), i.e.\ by white light. The latter is absorbed by the excited molecules and the amount as well as the frequency of the absorbed light is detected in a time-resolved fashion. During the relaxation, the transient absorption spectrum (given in the insets at the top) changes and the dynamics can be accurately mapped. Triplet states can also be detected in this way (e.g.\ by triplet-triplet absorption), but distinguishing singlet and triplet states is challenging. In general, assignment of transient absorption bands to specific processes is not an easy task and that is where theoretical simulations can help. It is in principle possible to calculate the excited-state absorption during the simulation of the dynamics. The bottleneck is that highly-excited states need to be computed and this can be difficult even for sophisticated methods, specially if accurate energetic differences are necessary in order to distinguish between all possible absorption pathways. 

As a final case, we mention also fluorescence experiments, where the fluorescence (excitation spectrum and fluorescence spectrum) of the excited-state population is detected time-dependently (Fig.~\ref{fig:22:exp}c). Although fluorescence can be simulated in principle, the timescales are often beyond what can be investigated in quantum or semiclassical on-the-fly dynamics calculations, and therefore theoretical studies are rare.

In a nutshell, the description of the complex processes underlying photorelaxation is very challenging and the corresponding simulation methods are subject to ongoing developments. However, keeping in mind all the pitfalls described above, extensive connections between theory and experiment are already possible today and a detailed picture of the photorelaxation dynamics of nucleic acids can be drawn.


\section{Photodynamics of Nucleobases}\label{sec:results}

In this section we shall discuss and analyze the photodynamics of each of the five DNA/RNA nucleobases in gas phase, starting with the purine bases A and G, followed by the pyrimidine bases C, T and U.

For each nucleobase we start briefly summarizing available experimental studies which report time scales and focus on the deactivation mechanism after light irradiation. These results will then be compared to the outcome of the theoretical studies. All relevant studies reporting excited-state dynamics simulations in the gas-phase are discussed in chronological order. An effort has been made to compare the proposed relaxation paths and the CoIns encountered in the simulations. Comparison between the studies is enabled by schematic overviews of the proposed relaxation mechanisms. As the reader will soon recognize, different methods provide different results, which sometimes complement to each other but sometimes conflict with each other. However, each dynamics method and each electronic structure method has different benefits and drawbacks, and it is often difficult to say \emph{which method is better}. As such, each simulation contributes with a small piece to the puzzle and advances a little further the interpretation of the experimental findings. An overall picture of the deactivation pathway for each nuclebase is summarized at the end of each subsection, assigning wherever possible the experimental time scales with the help of the molecular pathways obtained theoretically. All in all, this section shows the current state of knowledge of the deactivation processes happening in isolated photoexcited DNA/RNA nucleobases.


\subsection{Adenine}\label{ssec:results:ade}

\index{Adenine|(}
\index{Dynamics simulations!Adenine|(}
\index{Tautomers!Adenine}

\begin{figure}
  \sidecaption
  \includegraphics[scale=1]{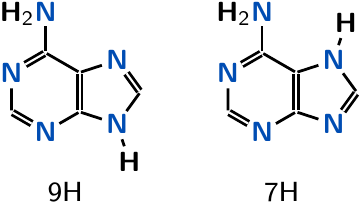}
  \caption{The most important tautomers of A. 9H A is the biologically relevant structure.}
  \label{fig:ade:tautomers}
\end{figure}

A is found in Watson-Crick pairs with T and U in DNA and RNA, respectively. Together with G it is a purine derivative and therefore exhibits qualitatively different features than the pyrimidine bases C, T or U. In water, A presents two different tautomers (see Fig.~\ref{fig:ade:tautomers}), the 7H and 9H-form, but 7H A is the minor fraction with 15--23\% of population~\cite{Cohen2003JACS, Chenon1975JACS, Dreyfus1975JACS, Gonella1983JACS, Holmen1997JPCA}.
In gas phase, 9H A is the only tautomer at biologically relevant temperatures~\cite{Bravaya2010JPCA}, although some studies also find 7H-A in gas phase when vaporizing at high temperatures~\cite{Plutzer2002PCCP}. Theoretically, only the dynamics of the 9H tautomer has been investigated, therefore by A we refer henceforth to this tautomer.

\subsubsection{Experimental timescales}\label{sssec:results:ade:exp}

Table~\ref{tab:ade:exp} summarizes the experiments that report time scales for the deactivation dynamics of A in gas phase, specifying which pump and probe wavelengths have been employed to that purpose. The papers are listed in chronological order and the time constants given in the corresponding publications have been classified \emph{by us}. As it will be seen, in all nucleobases different deactivation time scales have been found, typically spanning from fs to ps, or even to ns. In A, $\tau_1$ is assigned to constants below 100 fs, $\tau_2$ is considered for hundreds of fs, and $\tau_3$ is a ns time scale.

With this assignment in mind, all experiments show a fast deactivation ($\tau_2$) ranging from 0.5--2 ps. Additionally, most studies reported another, shorter transient below 100 fs, while Ullrich et al.~\cite{Ullrich2004JACS, Ullrich2004PCCP} even observed a long-lived component ($\tau_3$) on the ns time scale.

\begin{table}
  \centering
  \caption{Experimentally observed decay times of A.\label{tab:ade:exp}}
  \begin{tabular}{cccccllc}
    \hline
    \multicolumn{2}{c}{Setup}& \multicolumn{3}{c}{Time constants} &\multicolumn{2}{l}{Reference} &Year \\
    $\lambda_{\mathrm{pump}}$ (nm) &$\lambda_{\mathrm{probe}}$ (nm) & $\tau_1$ (fs) & $\tau_2$ (fs) & $\tau_3$ (ns) &\\
    \svhline
       267   &n$\times$800  &        &1000    &     &\cite{Kang2002JACS, Kang2003JCP}        & Kang et al.    & 2002,2003 \\
       250--277   &200       &$<$50   &750--2000 &1    &\cite{Ullrich2004JACS, Ullrich2004PCCP} & Ullrich et al. & 2004\\
       267   &2$\times$400  &100     &1000     &     &\cite{Canuel2005JCP}                    & Canuel et al.  & 2005\\
       267   &200           &40      &1200     &     &\cite{Satzger2006CPL, Satzger2006PNAS}  & Satzger et al. & 2006\\
       262   &n$\times$780  &100     &1140    &     &\cite{Kotur2012IJSTQE}                  & Kotur et al.   & 2012\\
    \hline
  \end{tabular}
\end{table}


\subsubsection{Deactivation mechanism}\label{sssec:results:ade:deact}

\begin{figure}
  \includegraphics[scale=1]{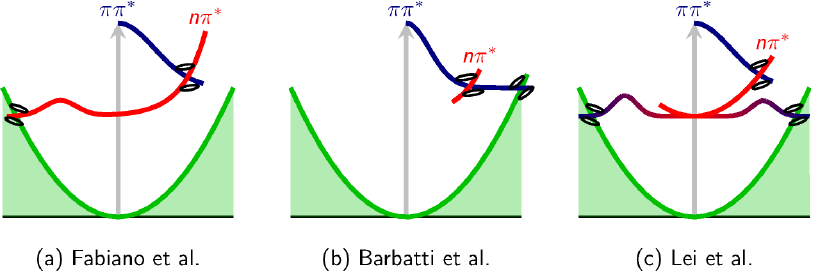}
  \caption{Schematic overview of the proposed relaxation mechanisms for A from the references: \textbf{a}~\cite{Fabiano2008JPCA}, \textbf{b}~\cite{Barbatti2008JACS,Barbatti2010PNAS,Barbatti2012JCP} and \textbf{c}~\cite{Lei2008JPCA} (left to right).}
  \label{fig:ade:deact}
  \index{Conical Intersections!Adenine}
\end{figure}

\begin{figure}
  \includegraphics[scale=1]{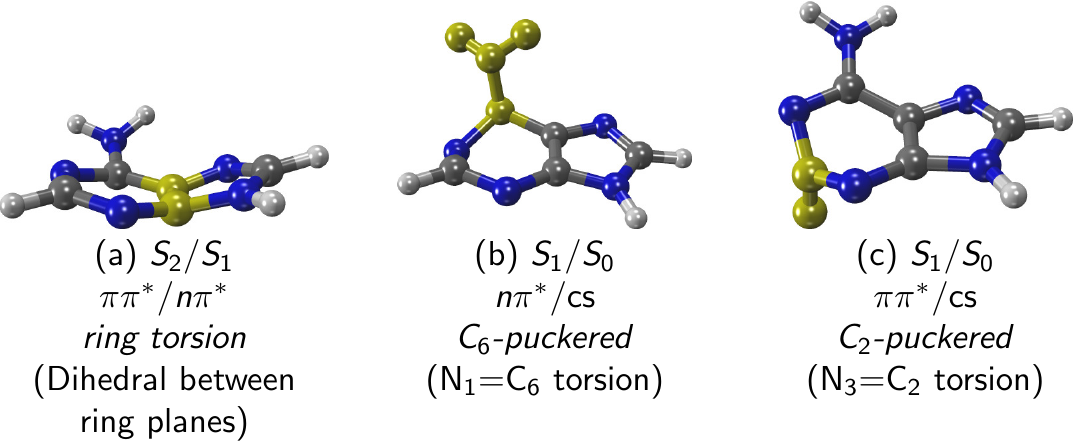}
  \caption{Geometries of important CoIns of A discussed in the text.
  The labels give the crossing states (adiabatic and state character) and the main geometrical feature, which are highlighted by golden atoms in the structures. }
  \label{fig:ade:coin}
\end{figure}

Table~\ref{tab:ade:theo} collects all the theoretical studies aimed at understanding the excited state deactivation dynamics of A in gas phase. Fig.~\ref{fig:ade:deact} depicts schematically each of the paths predicted by theoretical calculations. Note that these qualitative schemes collect several reaction coordinates in one dimension and are only intended for an at-first-glance comparison of the proposed relaxation pathways. For more detailed and precise information, the reader should consult the original reference. Colors indicate electronic state character, and are used consistently for all nucleobases. A color gradient indicates an adiabatic change of wavefunction character. If applicable, triplet states are given as dotted lines. Important geometries encountered by several studies are shown in Fig.~\ref{fig:ade:coin}. Atoms of characteristic geometrical features are given in gold. We note that in the dynamics the system does not hop exactly at the geometries shown, but at related geometries located on the same seam of intersection.

\begin{table}
  \centering
  \caption{Excited-state nuclear dynamics studies for isolated A in the gas phase.
  Time constants correspond to those given in the respective papers, classified as $\tau_1$ (below 120 fs) and $\tau_2$. 
  \label{tab:ade:theo}}
  \begin{tabular}{ccccllc}
    \hline
    \multicolumn{2}{c}{Methodology} &\multicolumn{2}{c}{Time constants}          &\multicolumn{2}{l}{Reference} &Year\\
    Dyn.    &El. Struct.            &$\tau_1$ (fs) &$\tau_2$ (fs)  &&&\\
    \svhline
     TSH   &OM2/MRCI                 & 15 & 560 &  \cite{Fabiano2008JPCA} & Fabiano et al. & 2008 \\
     TSH   &CASSCF(12,10)/MRCIS(6,4)$^a$ & 22 & 538 &  \cite{Barbatti2008JACS} & Barbatti et al. & 2008 \\
     MF    &DFTB                     &  & \hspace{.1cm}1050--1360 &  \cite{Lei2008JPCA} & Lei et al. & 2008 \\
     TSH   &TD-DFTB/B3LYP             & 120 & 11000 &  \cite{Mitric2009JPCA} & Mitric et al. & 2009 \\
     TSH   &CASSCF(12,10)/MRCIS(6,4)$^a$ &  & 440--770 &  \cite{Barbatti2010PNAS} & Barbatti et al. & 2010 \\
     TSH   &CASSCF(10,8)/MRCIS(6,4)  &  & 530 &  \cite{Barbatti2012JCP} & Barbatti et al. & 2012 \\
     TSH   &OM2/MRCI                 &  & 900 &  \cite{Barbatti2012JCP} & Barbatti et al. & 2012 \\
     TSH   &TD-DFT$^b$               &  & $\gg$1000 &  \cite{Barbatti2012JCP} & Barbatti et al. & 2012\\
    \hline
  \end{tabular}

  $^a$ Different analysis of the same data.
  $^b$ Functionals: CAM-B3LYP, B3LYP, BHLYP, M06-HF, PBE and PBE0.
\end{table}

The first excited-state dynamics study on gas phase A was conducted in 2008 by Fabiano et al.~\cite{Fabiano2008JPCA} using TSH, three electronic states and OM2/MRCI for the underlying electronic structure calculations. They identified processes on two timescales, which they assigned as $\tau_1$ and $\tau_2$. The fast time scale was described as the relaxation of the initially populated bright $\pi\pi^*$ state within 15 fs to a dark $n\pi^*$ state through the CoIn depicted qualitatively in  Fig.~\ref{fig:ade:coin}(a). The CoIn is chracterized by an angle of about 15$^\circ$ between the planes of the two rings. Once in the $n\pi^*$, the deactivation to the ground state takes 560 fs via the C$_6$-puckered ($n\pi^*/\mathrm{cs}$) CoIn, which features an out-of-plane distortion of the amino group (Fig.~\ref{fig:ade:coin}(b)). The overall path is depicted schematically in Fig.~\ref{fig:ade:deact}(a).

In the same year, a TSH study based on CASSCF(12,10)/MRCIS(6,4) electronic properties was conducted by Barbatti et al.~\cite{Barbatti2008JACS}. Their obtained time constants,$\tau_1=$22 fs and $\tau_2=$538 fs, are very similar to the ones of Fabiano et al.~\cite{Fabiano2008JPCA} but the details of the deactivation mechanisms differ slightly. As in Fabiano et al.~\cite{Fabiano2008JPCA}, the first constant arises from $S_2\rightarrow S_1$ decay and the second constant is connected to the $S_1\rightarrow S_0$ decay. However, and different from the previous study~\cite{Fabiano2008JPCA}, the CASSCF/MRCI-based dynamics employs the so-called C$_2$-puckered CoIn (Fig.~\ref{fig:ade:coin}(c)) as a major deactivation funnel. Note that, even though the authors reported $S_2\rightarrow S_1$ transitions, the system always stays in the $\pi\pi^*$ spectroscopic state.

Mean-field (MF) dynamics by Lei et al.~\cite{Lei2008JPCA} employing density functional-based tight binding (DFTB) observed a strong influence of the excitation energy on the relaxation path taken and hence on the relaxation times. Using an excitation energy of 5.0 eV, they observed that the C$_6$-puckered CoIn (Fig.~\ref{fig:ade:coin}b) is employed for relaxation. The excited-state lifetime in this case was 1050 fs (note that in this mean-field dynamics only a single trajectory is computed). On the other hand, excitation at 4.8 eV activates the channel through the C$_2$-puckered CoIn (Fig.~\ref{fig:ade:coin}c), with an excited-state lifetime of 1360 fs. In both cases the initial $\pi\pi^*$ state is reported to change to an intermediate $n\pi^*$ state before reverting to the $\pi\pi^*$ state and accessing the respective CoIn. Accordingly, the authors state that the final transition back to the ground state happens always from the $\pi\pi^*$ state (see Fig.~\ref{fig:ade:deact}c)---even though previous studies report the C$_6$-puckered CoIn to be of $n\pi^*/\mathrm{cs}$ character.

Mitri\'c et al.~\cite{Mitric2009JPCA} investigated the decay of photoexcited A  using TD-DFTB based dynamics. Here, for the isolated nucleobase a two-step relaxation process was observed, including a fast S$_2 \rightarrow$ S$_1$ transition with a time constant of 120 fs and a slow deactivation back to the ground state within 11 ps. Unfortunately, since their study mainly focused on the photodynamics of micro-solvated A, they did not report on the details of the relaxation path taken by gas-phase A.

In 2010, Barbatti et al.~\cite{Barbatti2010PNAS} published an overview article including all nucleobases, where the previously reported dynamics simulation~\cite{Barbatti2008JACS} for A was reanalyzed in terms of the initial excitation energy of the trajectories. Here, a monoexponential decay with a lifetime of about 770 fs was observed after excitation at the band origin (low energy), which is equivalent to an excitation wavelength of 267 nm. Increasing the excitation energy (250 nm) reduces the excited-state lifetime to 440 fs. Despite the different time constant, no differences in the mechanistic details of the relaxation were found for the different excitation energies.

The latest dynamics simulations on gas phase A were also published by Barbatti et al.~\cite{Barbatti2012JCP}, in a compendium paper which compares the results of dynamics simulations based on different electronic structure methods. The simulations showed that TD-DFT is not able to describe the ultrafast decay of excited A, even though six different functionals were employed. In none of the TD-DFT-based dynamics ground state relaxation was observed, except for very high initial energies. This is in strong contrast to the fast relaxation of A observed in the experiments. TD-DFT fails to describe the fast relaxation due to an overstabilization of planar distortions which gives a general underestimation of the puckering modes. On the contrary, OM2/MRCI and MRCIS dynamics show an ultrafast decay (900 or 530 fs, respectively) although via different deactivation channels that involve either puckering of the C$_6$ atom (OM2/MRCI) or the C$_2$ atom (MRCIS). Comparison of these results to reaction paths calculated at the CASPT2 and CC2 level of theory suggest that MRCIS underestimates the ease of C$_6$ puckering, whereas OM2/MRCI overestimates it~\cite{Barbatti2012JCP}.

\subsubsection{Final Discussion}\label{sssec:ade:sum}

Most of the dynamics studies of A in the gas phase assign the shortest time constant observed in the experiment ($\tau_1$) to $S_2\rightarrow S_1$ or $\pi\pi^*\rightarrow n\pi^*$ transitions occurring within the excited-state manifold. For the longer time constant $\tau_2$ the theoretically predicted values~\cite{Fabiano2008JPCA, Barbatti2008JACS, Barbatti2010PNAS, Barbatti2012JCP, Lei2008JPCA} corresponding to ground state relaxation are in the range 440--1120 fs, which is slightly shorter than the average of the experimental time constants. Nevertheless, given the good agreement, the assignment is plausible. The exception is provided by TD-DFT~\cite{Barbatti2012JCP} or TD-DFTB~\cite{Mitric2009JPCA} based TSH dynamics, which predict a significantly longer excited-state lifetime than observed experimentally, due to the limitations of these electronic structure methods to describe the excited-state PEHs.

Despite the fact that the time constants predicted by MRCIS and OM2/MRCI based dynamical studies~\cite{Fabiano2008JPCA, Barbatti2008JACS, Barbatti2010PNAS, Barbatti2012JCP} are similar, the obtained predominant relaxation pathways obtained are different. Depending on the level of theory, either the state character is preserved~\cite{Barbatti2008JACS,Barbatti2012JCP} leading to the C$_2$-puckered CoIn or it changes to $n\pi^*$~\cite{Fabiano2008JPCA, Barbatti2012JCP} leading to a decay via the C$_6$-puckered CoIn. It remains unclear which puckering motion is of major importance until on-the-fly dynamics simulations at a more reliable level of theory become possible.

\index{Adenine|)}
\index{Dynamics simulations!Adenine|)}


\subsection{Guanine}\label{ssec:results:gua}

\index{Guanine|(}
\index{Dynamics simulations!Guanine|(}
\index{Tautomers!Guanine}

\begin{figure}
  \includegraphics[scale=1]{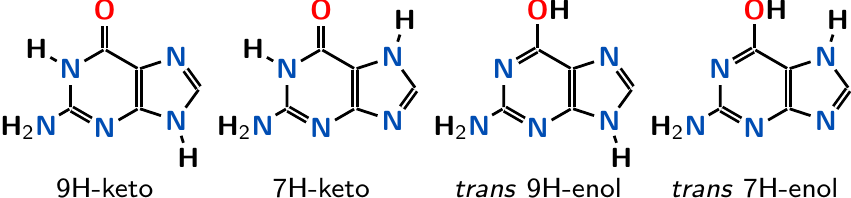}
  \caption{The most important tautomers of G. The biologically relevant form is the 9H-keto form.}
  \label{fig:gua:tau}
\end{figure}

G is another purine base; it is found in Watson-Crick pairs with C. G shows 36 different tautomers, from which the most important ones are shown in Fig.~\ref{fig:gua:tau}. Their experimental detection is very sensitive to the setup~\cite{Piuzzi2001CP, Mons2006JPCA, Yamazaki2008JPCA}. The most stable tautomers, 7H-keto-amino and 9H-keto-amino, show ultrafast decay. In DNA the 9H-keto-amino tautomer is dominant~\cite{Yamazaki2008JPCAa} and it shows a first absorption band at a maximum of 284 nm~\cite{Clark1965JPC}.

\subsubsection{Experimental observations}\label{sssec:results:gua:exp}

Table~\ref{tab:gua:exp} collects the experimentally decay times of G. Probably due to the large number of tautomers and difficulties associated, only a small number of experiments deal with the dynamics of photo-excited isolated G in gas phase. The pump-probe transient ionization experiments of Kang et al.~\cite{Kang2002JACS} find a monoexponential decay time of G after 267 nm excitation with a time constant of 800 fs in gas phase. Here, the cross-correlation of the pump and probe pulses of 400 fs (see~\cite{Kang2002CPL}) was probably to large to find the second short component, which was observed later on by Canuel et al.~\cite{Canuel2005JCP}. The latter authors reported $\tau_1=$148 fs and $\tau_2=$360 fs.

\begin{table}
  \centering
  \caption{Experimentally observed decay times of G.\label{tab:gua:exp}}
  \begin{tabular}{ccccllc}
    \hline
    \multicolumn{2}{c}{Setup}& \multicolumn{2}{c}{Time constants} &\multicolumn{2}{l}{Reference} &Year \\
    $\lambda_{\mathrm{pump}}$ (nm) &$\lambda_{\mathrm{probe}}$ (nm) & $\tau_1$ (fs) & $\tau_2$ (fs) &&&\\
    \svhline
    267 & n$\times$800 &   & 800 & \cite{Kang2002JACS} & Kang et al. & 2002 \\
    267 & 2$\times$400 & 148 & 360 & \cite{Canuel2005JCP} & Canuel et al. & 2005\\
    \hline
  \end{tabular}
\end{table}

\subsubsection{Deactivation mechanism}\label{sssec:results:gua:deact}

\begin{figure}
  \includegraphics[scale=1]{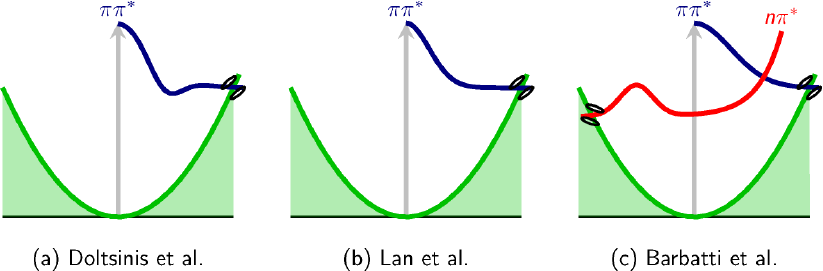}
  \caption{Schematic overview of the proposed relaxation mechanisms for 9H-keto G from~\textbf{a}~\cite{Doltsinis2008}, \textbf{b}~\cite{Lan2009CPC} and \textbf{c}~\cite{Barbatti2011JCP}. Note that different reaction coordinates can be implied in the one-dimensional pictures. }
  \label{fig:gua:deact}
\end{figure}

Table~\ref{tab:gua:theo} gives an overview of the different theoretical studies investigating gas phase dynamics of isolated G and Fig.~\ref{fig:gua:deact} shows the corresponding deactivation paths. 

\begin{table}
  \centering
  \caption{Excited-state nuclear dynamics studies for isolated G in the gas phase. The first column gives the tautomer (k: keto, e: enol). Time constants correspond to those given in the respective papers, classified as $\tau_1$ (below 100 fs) and $\tau_2$. 
  \label{tab:gua:theo}}
  \begin{tabular}{lccccllc}
    \hline
    G&\multicolumn{2}{c}{Methodology} &\multicolumn{2}{c}{Time constants}          &\multicolumn{2}{l}{Reference} &Year\\
    &Dyn.    &El. Struct.            &$\tau_1$ (fs) &$\tau_2$ (fs)  &&&\\
    \svhline
 7H-e &TSH-CP &ROKS/BLYP                &    &     &\cite{Langer2005CPC}    &Langer et al.    &2005\\
 9H-k &TSH-CP &ROKS/BLYP              &    &800  &\cite{Doltsinis2008}    &Doltsinis et al. &2008\\ 
 7H-k &TSH-CP &ROKS/BLYP              &    &1000 &\cite{Doltsinis2008}    &Doltsinis et al. &2008\\
 9H-k &TSH &OM2/MRCI(10,9)           &190 &400  &\cite{Lan2009CPC}       &Lan et al.       &2009\\
 9H-k &TSH &CASSCF(12,9)/MRCIS(10,7)$^a$ &    &280  &\cite{Barbatti2010PNAS} &Barbatti et al.  &2010\\
 9H-k &TSH &CASSCF(12,9)/MRCIS(10,7)$^a$ &    &224  &\cite{Barbatti2011JCP}  &Barbatti et al.  &2011\\
    \hline
  \end{tabular}

  $^a$ Same simulation, but different time constant reported.
\end{table}

The first dynamics study on G in the gas phase was reported by Langer et al.~\cite{Langer2005CPC} in 2005. The method used was a TSH approach coupled to Car-Parinello dynamics (TSH-CP), based on PEHs from restricted open-shell Kohn-Sham (ROKS) with the BLYP functional. They investigated the possibility of O--H dissociation in the excited state dynamics of 7H-enol G. Since this reaction involves a rather high barrier, constrained dynamics in order to sample this barrier was carried out. It was found that in gas phase the barrier is about 0.54 eV and thus is likely only relevant at high excitation energies. The abstraction of the hydrogen atom was accompanied by a non-adiabatic change to the $S_0$ state through an $S_1/S_0$ CoIn.

\begin{figure}
  \includegraphics[scale=1]{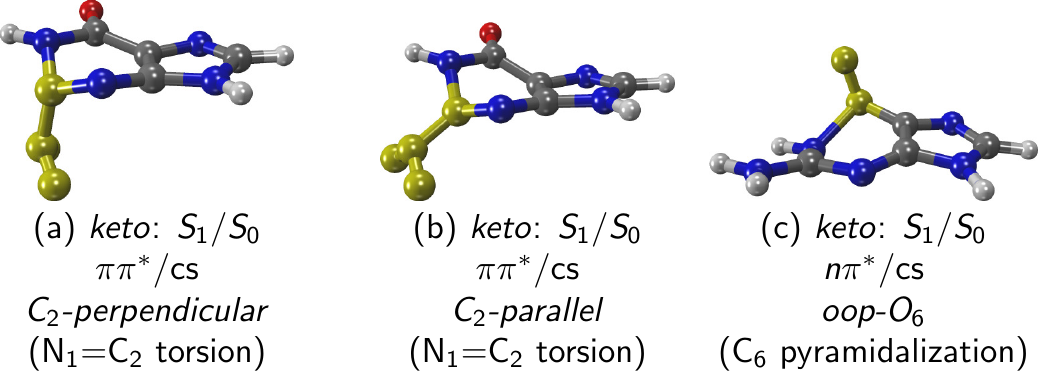}
  \caption{Geometries of important CoIns of G discussed in the text. The labels give the crossing states (adiabatic and state character) and the main geometrical feature. }
  \label{fig:gua:coin}
  \index{Conical Intersections!Guanine}
\end{figure}

In 2008, Doltsinis et al.~\cite{Langer2006,Doltsinis2008} for the first time reported excited-state dynamics of G focused on understanding the photodeactivation. Using the same method as in~\cite{Langer2005CPC}, they reported a monoexponential decay for 9H-keto G with a time constant of 800 fs. Here, the system first relaxes to the $S_1$ ($\pi\pi^*$) minimum, before reaching an $S_1/S_0$ CoIn, as shown schematically in Fig.~\ref{fig:gua:deact}a. The CoIn is characterized by C$_2$ puckering and an N$_2$--C$_2$--C$_4$--C$_5$ dihedral angle of 97$^\circ$ (see Fig.~\ref{fig:gua:coin}a). For the 7H-keto tautomer the $S_1$ minimum is very similar to the ground state minimum geometry. The decay dynamics are mostly driven by distortions of the imidazole ring and out-of-plane motions of the oxygen atom, resembling the oop-O$_6$ CoIn which is given in Fig.~\ref{fig:gua:coin}c for the 9H form. The excited-state population decays monoexponentially to the ground state with a time constant of 1000 fs.

Simulations based on TSH and the semi-empirical OM2/MRCI electronic structure method by Lan et al.~\cite{Lan2009CPC} find two different deactivation channels for 9H-keto G. The first channel involves the C$_2$-parallel CoIn given in Fig.~\ref{fig:gua:coin}b, which is stated to be responsible for a fast decay with a time constant of 190 fs. The second channel leads through the C$_2$-perpendicular CoIn, depicted in Fig.~\ref{fig:gua:coin}a. A 400 fs time constant was reported for this pathway. The two CoIns are distinguished by the angle between the C$_2$--amino bond and the ring plane. This angle is much larger for the C$_2$-parallel CoIn than for the C$_2$-perpendicular CoIn. Note that in this study the relaxation mechanism involved only the excited $\pi\pi^*$ state and the ground state (see Fig.~\ref{fig:gua:deact}b), even though an $n\pi^*$ state was included in the calculations.

In 2010, Barbatti et al.~\cite{Barbatti2010PNAS} published a TSH dynamics study based on the correlated CASSCF(12,9)/MRCIS(10,7) level of theory. They reported monoexponential decay for 9H-keto G with a time constant of 280 fs, caused by a single and direct decay via the two C$_2$-puckered CoIns (Fig.~\ref{fig:gua:coin}a and b). The authors stated that both CoIns involve an ethylene-like twisting of the N$_1$=C$_2$ bond, can be reached barriereless from the FC region and connect the $\pi\pi^*$ state to the ground state. A more detailed analysis of the same simulations~\cite{Barbatti2011JCP} revealed a slightly shorter time constant of 224 fs, which is a composite of a 97 fs delay and a 127 fs monoexponential decay of the excited-state population. The main route of deactivation follows the $\pi\pi^*$ state towards the C$_2$-parallel and C$_2$-perpendicular CoIns (Fig.~\ref{fig:gua:coin}a and b), which together account for 95\% of all hops. Only 5\% of the trajectories showed instead a transition to the $n\pi^*$ state, followed by out-of-plane motion of the oxygen atom, leading to the $S_1/S_0$ oop-O$_6$ CoIn (Fig.~\ref{fig:gua:coin}c).

\subsubsection{Final Discussion}\label{sssec:gua:sum}

For G, less experimental and theoretical studies focused on the gas-phase excited-state dynamics are available compared to the rest of the nucleobases. 

The existing literature identifies G as the nucleobases showing the fastest relaxation to the ground state. Experimentally, a sub-ps decay is reported, which is also reproduced by all dynamics simulations. While the ROKS-based study~\cite{Doltsinis2008} is consistent with the experimental transients of Kang et al.~\cite{Kang2002JACS}, the later simulations employing OM2/MRCI or CASSCF/MRCIS show good agreement with the time constants of Canuel et al.~\cite{Canuel2005JCP}. All studies agreed on the dominance of the direct $\pi\pi^*\rightarrow S_0$ pathway, accessing the C$_2$-puckered CoIns. Relaxation involving the $n\pi^*$ state is likely to be of minor importance.

Based on the simulations of Doltsinis et al.~\cite{Doltsinis2008}, the 9H and 7H tautomers of G are expected to show decay on a very similar time scale. Nevertheless, in order to clarify the involvement of the 7H tautomer in the photodynamics of G additional dynamics simulations based on reliable electronic structure methods are necessary.

\index{Dynamics simulations!Guanine|)}
\index{Guanine|)}


\subsection{Cytosine}\label{ssec:results:cyt}

\index{Cytosine|(}
\index{Dynamics simulations!Cytosine|(}
\index{Tautomers!Cytosine}

\begin{figure}
  \includegraphics[width=\textwidth]{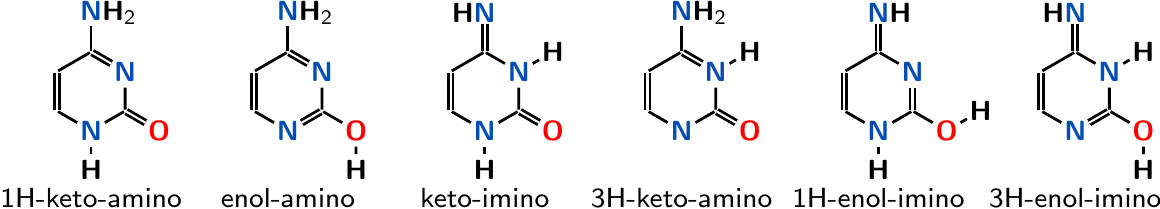}
  \caption{The six possible tautomers of C. The 1H-keto-amino form is found in DNA.}
  \label{fig:cyt:tautomer}
\end{figure}

C is a pyrimidine derivative, in contrast to A and G. In Watson-Crick pairs, C is hydrogen-bonded to G. C can exist in different tautomers, which are shown in Fig.~\ref{fig:cyt:tautomer}. The relative stability of the tautomers depends on the environmental conditions. While in DNA and in aqueous solution C only exists in the 1-H-keto-amino form (henceforth denoted as keto), in gas phase other tautomers have been detected. A microwave spectroscopic study by Brown et al.~\cite{Brown1989JACS} found a tautomer ratio of 0.44:0.44:0.12, for the keto, the enol-amino (enol) and the keto-imino (imino) tautomers, respectively. Tautomers other than keto, enol and imino are very high in energy and have not been detected in the experiments. They are not discussed in the following. Two other matrix isolation infrared studies obtained ratios of 0.32:0.65:0.03~\cite{Szczesniak1988JACS} and 0.22:0.70:0.08~\cite{Bazso2011PCCP}. Since the absorption spectra of the three tautomers overlap, it is expected that in gas-phase experimental studies the signals of these three tautomers are measured simultaneously, giving rise to several time constants or effective time constants and making the assignment of time scales exceptionally difficult. For this reason, theoretical studies can be particularly helpful to assist the interpretation of the available time-resolved spectra. However, as it will be discussed below, most theoretical gas phase studies have been focused only on the keto form, most likely because it is the biologically relevant tautomer. In the following, if no tautomer form is explicitly given, the keto form is implicitly assumed. 


\subsubsection{Experimental Observations}\label{sssec:results:cyt:exp}

\begin{table}
  \centering
  \caption{Experimentally observed decay times of C\label{tab:cyt:exp}. }
  \begin{tabular}{cccccllc}
    \hline
    \multicolumn{2}{c}{Setup}& \multicolumn{3}{c}{Time constants} &\multicolumn{2}{l}{Reference} &Year \\
    $\lambda_{\mathrm{pump}}$ (nm) &$\lambda_{\mathrm{probe}}$ (nm) & $\tau_1$ (fs) & $\tau_2$ (fs) & $\tau_3$ (ps) &\\
    \svhline
    267       &$n\times$800   &        &            &3.2        &\cite{Kang2002JACS}    &Kang et al.    &2002\\
    250       &200            &$<$50   &820         &3.2        &\cite{Ullrich2004PCCP} &Ullrich et al. &2004\\
    267       &2$\times$400   &160     &            &1.86       &\cite{Canuel2005JCP}   &Canuel et al.  &2005\\
    260--290  &3$\times$800   &200--100&3800--1100  &up to 150  &\cite{Kosma2009JACS}   &Kosma et al.   &2009\\
    260--300  &3$\times$800   &        &200--1200   &2.7-45     &\cite{Ho2011JPCA}      &Ho et al.      &2011\\
    262       &$n\times$780   &50      &240         &2.36       &\cite{Kotur2012IJSTQE} &Kotur et al.   &2012\\
    \hline
  \end{tabular}
\end{table}

There are a number of experimental studies that have measured the excited-state lifetimes of C in the gas phase. Table~\ref{tab:cyt:exp} lists the exponential decay time constants reported in these studies. In most experiments a single pump wavelength~\cite{Kang2002JACS,Ullrich2004PCCP,Canuel2005JCP,Kotur2012IJSTQE} has been used. Kosma et al.~\cite{Kosma2009JACS} and Ho et al.~\cite{Ho2011JPCA} employed arrays of excitation wavelengths ranging from 260 nm to 300 nm.

Generally, most of these papers report a time constant of about 1 ps ($\tau_2$) and a longer time constant of several ps ($\tau_3$). Interestingly, the shorter constant is present at all excitation wavelengths, but the longer time constant strongly increases with increasing excitation wavelength and vanishes for $\lambda_{\mathrm{pump}}>$290 nm. Besides these two time constants, some studies resolved another, very short decay constant ($\tau_1$).

In order to disentangle the deactivation pathways of each of the tautomers, Ho et al.~\cite{Ho2011JPCA} also carried out experiments for 1-methyl-C and 5-fluoro-C. In the former molecule, which cannot tautomerize to the enol form, they found essentially the same values for $\tau_2$ as in C but the longer time scale $\tau_3$ disappears. In 5-fluoro-C, (which assumes the enol form), the dynamics was completely dominated by $\tau_3$.


\subsubsection{Deactivation Mechanism}\label{sssec:results:cyt:deact}

\begin{table}
  \centering
  \caption{Excited-state nuclear dynamics studies for isolated C in the gas phase.
  The first column gives the tautomer (k: keto, e: enol, i: imino). Time constants correspond to those given in the respective papers, classified as $\tau_1$ below 100 fs, $\tau_2$ below 1 ps, and $\tau_3$ above 1 ps. A checkmark indicates that the authors discussed processes on these timescales without giving explicit values.
  \label{tab:cyt:theo}}.
  \begin{tabular}{lcccccllc}
    \hline
    C              &\multicolumn{2}{c}{Methodology} &\multicolumn{3}{c}{Time constants} &\multicolumn{2}{l}{Reference} &Year\\
                        &Dyn.    &El. Struct.       &$\tau_1$ (fs) &$\tau_2$ (fs) &$\tau_3$ (ps) &&&\\
    \svhline
    k &TSH-CP    &ROKS/BLYP             &       &700     &       &\cite{Doltsinis2008}                     &Doltsinis et al.            &2008\\
    k &FMS       &CASSCF(2,2)           &$\surd$&        &       &\cite{Hudock2008CPC}                     &Hudock et al.               &2008\\
    k &TSH       &OM2/MRCI              &40     &370     &       &\cite{Lan2009JPCB}                       &Lan et al.                  &2009\\
    k &TSH       &AM1/CI(2,2)           &       &        &       &\cite{Alexandrova2010JPCB}               &Alexandrova et al.          &2010\\
    k &TSH       &CASSCF(12,9)          &$\surd$&$\surd$ &       &\cite{Gonzalez-vazquez2010CPC}           &Gonz\'alez-V\'azquez et al. &2010\\
    k &TSH       &CASSCF(14,10)         &9      &527     &3.08   &\cite{Barbatti2010PNAS,Barbatti2011PCCP} &Barbatti et al.             &2010,2011\\
    k &\textsc{Sharc} &CASSCF(12,9)     &$\surd$&        &       &\cite{Richter2012JPCL}                   &Richter et al.              &2012\\
    k &TSH       &CASSCF(12,9)          &       &$\surd$ &       &\cite{Nakayama2013PCCP}                  &Nakayama et al.             &2013\\
    e &TSH       &CASSCF(10,8)          &       &        &$\surd$&\cite{Nakayama2013PCCP}                  &Nakayama et al.             &2013\\
    i &TSH       &CASSCF(12,9)          &$\surd$&        &       &\cite{Nakayama2013PCCP}                  &Nakayama et al.             &2013\\
    k &\textsc{Sharc} &CASSCF(12,9)     &7      &270     &       &\cite{Mai2013CPC}                        &Mai et al.                  &2013\\
    e &\textsc{Sharc} &CASSCF(12,9)     &40     &        &1.9    &\cite{Mai2013CPC}                        &Mai et al.                  &2013\\
    \hline
  \end{tabular}
\end{table}

\begin{figure}
  \includegraphics[scale=1]{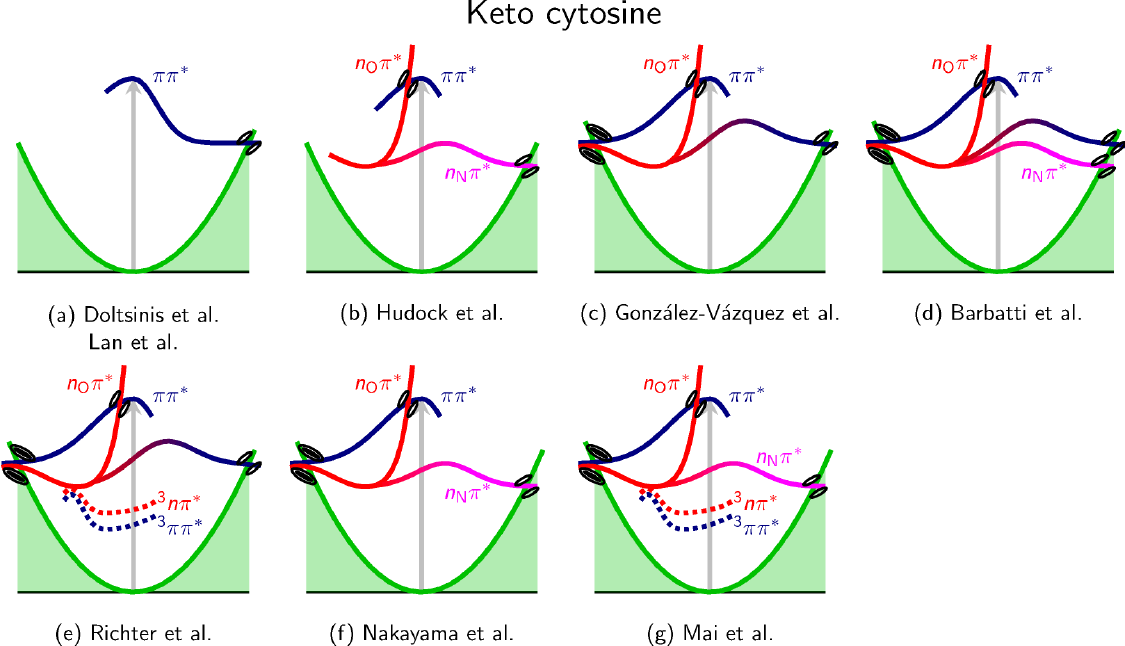}
  \caption{Schematic overview of the proposed relaxation mechanisms for keto C from
  \textbf{a}~\cite{Doltsinis2008, Lan2009JPCB},
  \textbf{b}~\cite{Hudock2008CPC},
  \textbf{c}~\cite{Gonzalez-vazquez2010CPC},
  \textbf{d}~\cite{Barbatti2011PCCP},
  \textbf{e}~\cite{Richter2012JPCL},
  \textbf{f}~\cite{Nakayama2013PCCP} and
  \textbf{f}~\cite{Mai2013CPC}.
  Note that different reaction coordinates can be implied in the one-dimensional picture.}
  \label{fig:cyt:deact}
\end{figure}

Table~\ref{tab:cyt:theo} reports a comprehensive list of all the excited-state dynamical studies performed in isolated C in the gas phase and Fig.~\ref{fig:cyt:deact} depicts schematically all the proposed mechanisms. They will be discussed in chronological order in the following.

\begin{figure}
  \includegraphics[scale=1]{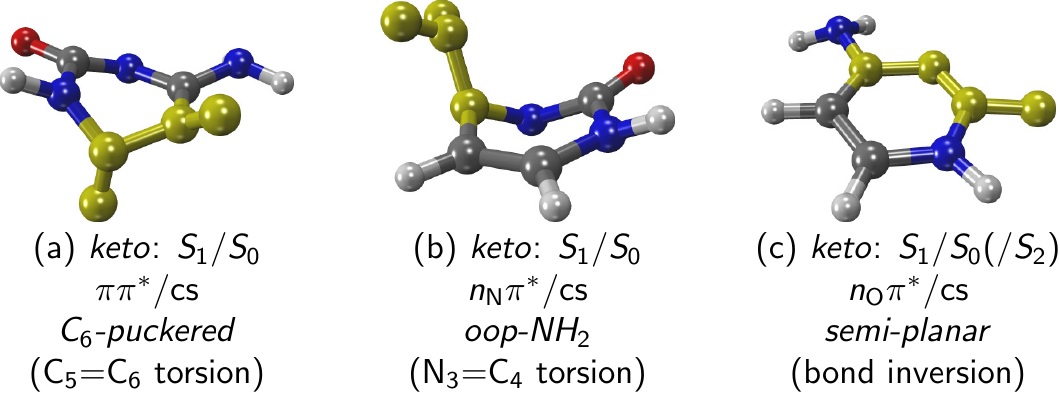}
  \caption{Geometries of important CoIns of keto C discussed in the text. The labels give the crossing states (adiabatic and state character) and the main geometrical feature. The geometry of the semi-planar CoIn in \textbf{c} is similar to the three-state CoIn mentioned in the text.}
  \label{fig:cyt:coin}
  \index{Conical Intersections!Cytosine}
\end{figure}

The first dynamics simulation for keto C was reported by the group of Doltsinis~\cite{Doltsinis2008} in 2008, who employed the TSH methodology coupled to Car-Parinello dynamics (TSH-CP). The underlying electronic structure calculations were performed with restricted open-shell Kohn-Sham (ROKS)~\cite{Doltsinis2008} and the BLYP functional and included only the ground and first singlet excited state, which is of $\pi\pi^*$ character. The authors observed a mono-exponential 700 fs lifetime of the excited state, in good agreement with the 820 fs transient observed experimentally by Ullrich et al.~\cite{Ullrich2004PCCP}. For the $S_1$ state no change of character was observed, as the $\pi\pi^*$ state was the only excited state involved in the calculations. The qualitative picture of the dynamics is summarized in Fig.~\ref{fig:cyt:deact}(a). The main relaxation mechanism involved variations of the C$_5$=C$_6$ bond length and the H$_5$--C$_5$=C$_6$--H$_6$ dihedral~\cite{Langer2006}, as shown for the so-called C$_6$-puckered CoIn in Fig.~\ref{fig:cyt:coin}(a).

In the same year, Hudock et al.~\cite{Hudock2008CPC} performed an AIMS study in keto C based on CASSCF(2,2) calculations, also restricted to the ground and the first excited state. This two-state treatment does not allow to describe three-state CoIns, which had been suggested by Matsika and coworkers using quantum chemical calculations~\cite{Kistler2008JCP}, but based on~\cite{Blancafort2004JPCA}, it was argued~\cite{Hudock2008CPC} that these CoIns are energetically inaccessible and therefore two states should suffice to describe the dynamics of C. As a first step in the dynamics, an adiabatic change of character from $\pi\pi^*$ to either $n_\mathrm{N}\pi^*$ or $n_\mathrm{O}\pi^*$ is predicted within the first 100 fs. Most of the population (65\%) relaxed within 1 ps through a CoIn where $n_\mathrm{N}\pi^*$ and ground state cross. This CoIn features a strong out-of-plane distortion of the amino group, hence it is denoted as oop-NH$_2$ CoIn in the following (see Fig.~\ref{fig:cyt:coin}(b)). A smaller number of basis functions also relaxed through the $C_6$-puckered CoIn (Fig.~\ref{fig:cyt:coin}(a)) or the semi-planar CoIn (Fig.~\ref{fig:cyt:coin}(c)). The latter CoIn connects the $n_\mathrm{O}\pi^*$ with the ground state and shows a bond inversion of the C$_4$=N$_3$--C$_2$=O moeity, with stretching of the C$_4$=N$_3$ and C$_2$=O bonds and compression of the N$_3$--C$_2$ bond. Despite the fact that this study only considered two electronic states, it already reveals that keto C can follow a number of different pathways after excitation. Fig.~\ref{fig:cyt:deact}(b) summarizes the most prevalent ones: the relaxation from $\pi\pi^*$ to the $n_\mathrm{O}\pi^*$ state, from which the system changes to the $n_\mathrm{N}\pi^*$ character adiabatically (indicated by a color gradient) and later decays to the closed-shell ground state through the $n_\mathrm{N}\pi^*/\mathrm{cs}$ CoIn.

One year later, Lan et al.~\cite{Lan2009JPCB} published a TSH study based on the semiempirical OM2/MRCI electronic structure method, including three electronic states---ground state, $\pi\pi^*$ and $n\pi^*$. For most trajectories the initial state was $S_1$; however, since at the Franck-Condon region both $\pi\pi^*$ and $n\pi^*$ states are very close, for some geometries the bright $\pi\pi^*$ state corresponded to $S_2$, as explained in Sect.~\ref{sec:sim_exp} They obtained two time constants, $\tau_1=$40 fs for the $S_2\rightarrow S_1$ decay and $\tau_2=$370 for the $S_1\rightarrow S_0$ relaxation. Particularly, the second time constant is too short---as the authors state---compared to the experimental value of 820 fs~\cite{Ullrich2004PCCP}, but they ascribe the error to the approximations made in the calculations. Interestingly, even though the simulations included both the $\pi\pi^*$ and $n\pi^*$ states, the system exclusively relaxes to the ground state through the $\pi\pi^*/\mathrm{cs}$ C$_6$-puckered CoIn (Fig.~\ref{fig:cyt:coin}(a)), and the $n\pi^*$ state was not reported to play a major role in the dynamics. Overall, this is a very similar mechanism as that proposed by Doltsinis et al.~\cite{Doltsinis2008} (see Fig.~\ref{fig:cyt:deact} (a)), even though the time constants of Doltsinis et al.\ and Lan et al.\ differ by a factor of two.

Alexandrova and coworkers~\cite{Alexandrova2010JPCB} performed TSH simulations based on semiempirical AM1/MRCI(2,2) calculations on several nucleosides and DNA fragments, and also on isolated keto C. While they give time constants of 60--120 fs for cytidine, unfortunately, no time constants are reported for C. In their simulations, most of the trajectories relaxed with an out-of-plane motion of the C$_4$ atom, which might correspond to the oop-NH$_2$ CoIn shown in Fig.~\ref{fig:cyt:coin}(b). They also observed a very large stretch of the N$_1$--C$_2$ bond at hopping geometries.

Later TSH studies on C were done using CASSCF electronic structure calculations with relatively large active spaces. The first of these studies was carried out by our group~\cite{Gonzalez-vazquez2010CPC} using CASSCF(12,9) PEHs, four singlet states, and a total propagation time of 200 fs. After this time, it was found that already 35\% of the trajectories of keto C returned to the ground state, while 10\% were trapped in the $S_2$. The former pathway was assigned as responsible of the fast ($\tau_1$) experimentally observed transients and the latter to the slower transients ($\tau_2$), although no decay constants were calculated because of the short propagation times. These calculations were superseded in later publications~\cite{Richter2012JPCL,Mai2013CPC}, but very importantly in this work it was shown that the three-state CoIn reported much earlier by static calculations~\cite{Blancafort2004JPCA, Kistler2008JCP} is a key ingredient in the deactivation pathway of C. At this CoIn the $\pi\pi^*$, $n_\mathrm{O}\pi^*$ and ground state intersect. The geometry of this CoIn is comparable to the semi-planar $n_\mathrm{O}\pi^*/\mathrm{cs}$ CoIn (Fig.~\ref{fig:cyt:coin}(c)), with additional pyramidalization of the C$_6$ atom~\cite{Kistler2008JCP}. The deactivation channel involving the semi-planar CoIn and the nearby three-state CoIn is shown qualitatively on the left side of Fig.~\ref{fig:cyt:deact}(c). The simulations~\cite{Gonzalez-vazquez2010CPC} also found a decay pathway through the C$_6$-puckered CoIn (Fig.~\ref{fig:cyt:coin}(a)), where deactivation was slightly slower than through the semi-planar CoIn. The path to this CoIn is shown on the right side of Fig.~\ref{fig:cyt:deact}(c).

The later work of Barbatti et al.~\cite{Barbatti2010PNAS,Barbatti2011PCCP} on keto C based on TSH simulations built on CASSCF(14,10) calculations, reports three time constants, $\tau_1=$9 fs, $\tau_2=$527 fs and $\tau_3=$3.08 ps, as a result of a triexponential fit of the ground state population of the first 1.2 ps. The time constants are in quite good agreement with those reported experimentally in~\cite{Kang2002JACS, Ullrich2004PCCP}; however, they represent the average over several different relaxation pathways. Similar to Gonz\'alez-V\'azquez et al.~\cite{Gonzalez-vazquez2010CPC}, the authors of~\cite{Barbatti2010PNAS,Barbatti2011PCCP} find that the semi-planar CoIn (Fig.~\ref{fig:cyt:coin} (c)) is the most relevant one. During the first 10 fs, 16\% of the ensemble returned to the ground state through this CoIn. Also for later times, the main decay pathway involves the semi-planar CoIn, as indicated on the left side of Fig.~\ref{fig:cyt:deact}(d). Additional minor pathways involve either a switch to the $n_\mathrm{N}\pi^*$ state and a relaxation through the oop-NH$_2$ CoIn (Fig.~\ref{fig:cyt:coin} (b)) or a recrossing to the $\pi\pi^*$ state with subsequent decay through the C$_6$-puckered CoIn (Fig.~\ref{fig:cyt:coin} (a)). Both paths are represented on the right side of Fig.~\ref{fig:cyt:deact} (d).

The first study to include triplet states in the dynamics was performed in our group~\cite{Richter2012JPCL} in 2012 using the \textsc{Sharc} methodology. The simulations were done using CASSCF(12,9), as in~\cite{Gonzalez-vazquez2010CPC}, but propagating for 1 ps. A detailed analysis of the multiple pathways taking place among the adiabatic states is done---most of them with time constants below 100 fs. After 1 ps, 90\% of the ensemble returned to the ground state, as in the simulations of Barbatti et al.~\cite{Barbatti2011PCCP}. The reported hopping geometries also showed similarities with the C$_6$-puckered and semi-planar CoIns (Fig.~\ref{fig:cyt:coin} (a) and (c)), but not with the oop-NH$_2$ CoIn. In this sense, the radiationless IC pathways involved the $\pi\pi^*$ and $n_\mathrm{O}\pi^*$ states, as in our previous study~\cite{Gonzalez-vazquez2010CPC}. Accordingly, Fig.~\ref{fig:cyt:deact}(e) is the same for the singlet deactivation, as Fig.~\ref{fig:cyt:deact}(c). However, and very interestingly, the remaining 10\% of the ensemble underwent ISC on a sub-ps timescale. The doorway state is the $S_1$ ($n_\mathrm{O}\pi^*$), which interacts with the $^3\pi\pi^*$ and $^3n\pi^*$ states. According to the CASSCF on-the-fly calculations ISC was found primarily between the $S_1$ and $T_2$ states. Subsequent triplet IC leads to the $T_1$ minimum, where the trajectories get trapped for the remainder of the simulations. This is illustrated also in Fig.~\ref{fig:cyt:deact}(e).

Up to this point, all the papers discussed have been concerned only with the keto form of C. However, as mentioned above, in gas phase several tautomeric forms coexists, with the keto, enol and imino tautomers being the most important ones. Thus, a comprehensive analysis of the reported experiments requires to study the dynamics of the enol and imino tautomers as well. This fact has been recognized in 2013, where two papers including several tautomers have been published~\cite{Nakayama2013PCCP,Mai2013CPC}.

\begin{figure}
  \includegraphics[scale=1]{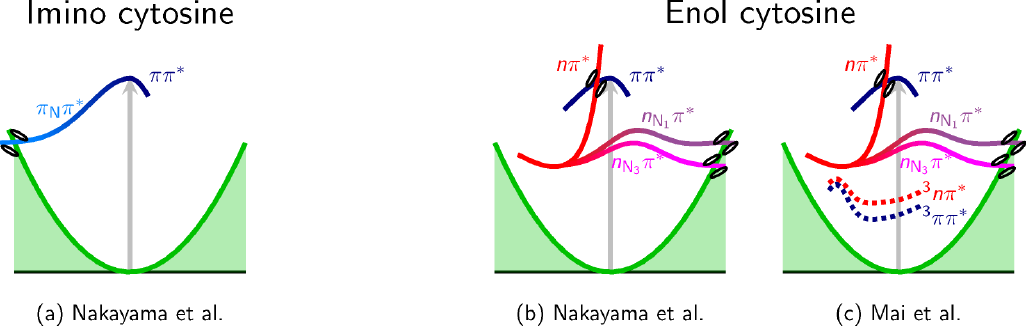}
  \caption{Qualitative overview over the proposed relaxation mechanisms for enol- and imino-C from the studies
  \textbf{a} and \textbf{b} of Nakayama et al.~\cite{Nakayama2013PCCP} and \textbf{c} of Mai et al.~\cite{Mai2013CPC}.}
  \label{fig:cyt:deact2}
\end{figure}

\begin{figure}
  \includegraphics[scale=1]{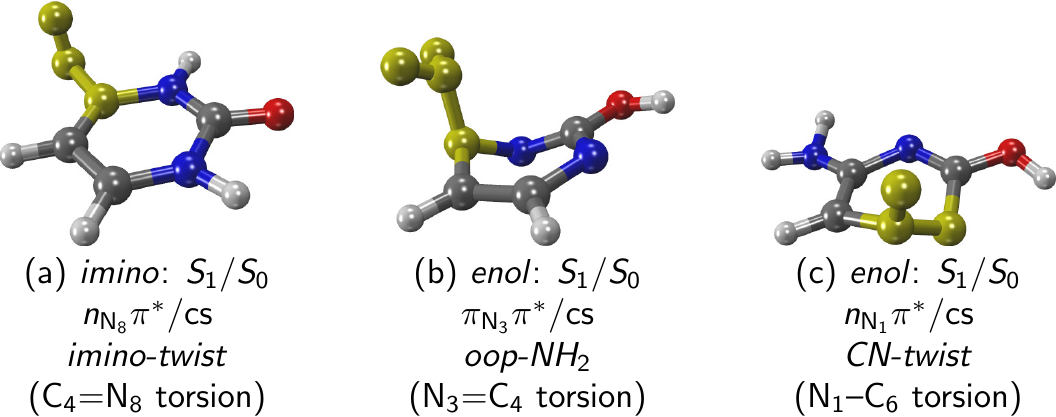}
  \caption{Geometries of important CoIns of enol and imino C discussed in the text.
  The labels give the tautomer, crossing states (adiabatic and state character) and the main geometrical feature.}
  \label{fig:cyt:coin2}
  \index{Conical Intersections!Cytosine}
\end{figure}

Nakayama et al.~\cite{Nakayama2013PCCP} performed TSH simulations on the keto, enol and imino forms of C, taking into account only singlet states. For keto and imino, they used CASSCF(12,9) electronic structure calculations, while for the enol tautomer, CASSCF(10,8) was employed. They found that the imino tautomer clearly shows the fastest decay of the three tautomers, with all trajectories returning to the ground state in less than 200 fs. The keto form showed a slower relaxation---85\% of the ensemble relaxed within 1 ps, which is comparable to the results obtained by Barbatti et al.~\cite{Barbatti2011PCCP} and Richter et al.~\cite{Richter2012JPCL}. The slowest decay was observed for the enol form, where only 10\% of the population relaxed in the first ps. The assignment of intermediate time constants to keto and long time constants to enol is consistent with the experimental results of Ho et al.~\cite{Ho2011JPCA}. Nakayama et al.~\cite{Nakayama2013PCCP} showed that each tautomer relaxes through different pathways. The fast relaxation of the imino form is mediated by the rotation of the imino-hydrogen, as shown in Fig.~\ref{fig:cyt:coin2} (a). This process is accompanied by a change of character to $\pi_{\mathrm{N}_8}\pi^*$ (see Fig.~\ref{fig:cyt:deact2} (a)). Since the hydrogen atom is very light and there is no barrier, the relaxation is extremely efficient. In keto C, the authors find a strong predominance of the route involving the semi-planar CoIn (Fig.~\ref{fig:cyt:coin} (c)), in agreement with similar previous CASSCF studies~\cite{Gonzalez-vazquez2010CPC, Barbatti2011PCCP, Richter2012JPCL}. They also found a small number of trajectories relaxing via $n_\mathrm{N}\pi^*$ through the oop-NH$_2$ CoIn, but no relaxation via $\pi\pi^*$ through the C$_6$-puckered CoIn (see Fig.~\ref{fig:cyt:deact} (f)). For the enol form, they found a fast relaxation from the FC region ($\pi\pi^*$) to the $n\pi^*$ minimum. A small number of trajectories relaxed through two CoIns: the first one is also of oop-NH$_2$ type (Fig.~\ref{fig:cyt:coin2} (b)) and the second one is termed CN-twist CoIn since it involves a puckering of N$_1$ and C$_6$ (Fig.~\ref{fig:cyt:coin2} (c)).
The initial relaxation to the $n\pi^*$ minimum and the two decay pathways are depicted in Fig.~\ref{fig:cyt:deact2} (b).

The most recent study including different C tautomers has been conducted in our group~\cite{Mai2013CPC}, including both singlets and triplet states by means of the \textsc{Sharc} method. The simulations considered the keto and the enol forms of C. For the keto tautomer time constants of $\tau_1=$7 fs and $\tau_2=$270 fs were obtained. These time constants included both decay paths to the $S_0$ ground state as well as the deactivation into the $T_1$ minimum, explaining the smaller time constants compared to the other CASSCF dynamics simulations~\cite{Gonzalez-vazquez2010CPC, Barbatti2011PCCP}. For enol a long time constant of 1900 fs was found, which can be assigned to $\tau_3$. In the enol form, ISC was observed with very low efficiency and hence no time constant related to ISC was stated.
The IC pathways of keto C were basically the same as in the study by Nakayama et al.~\cite{Nakayama2013PCCP}. The most important CoIn was the semi-planar one (Fig.~\ref{fig:cyt:coin} (c)), but the oop-NH$_2$ CoIn (Fig.~\ref{fig:cyt:coin} (b)) also played a significant role. In contrast, the C$_6$-puckered CoIn was not important in the dynamics. The ISC pathways are in general agreement with the ones previously calculated by Richter et al.~\cite{Richter2012JPCL}. The $S_1$ ($n_\mathrm{O}\pi^*$) state is the precursor of the triplet states and ISC takes place mainly to the $T_2$ ($^3n\pi^*$) state, with higher efficiency when $T_1$ and $T_2$ are very close and mix. This mixing increases the otherwise small spin-orbit couplings between $S_1$ and $T_2$. The IC and ISC pathways are shown schematically in Fig.~\ref{fig:cyt:deact} (g).
For enol, the dynamics observed is similar to that reported by Nakayama et al.~\cite{Nakayama2013PCCP}. However, Mai et al.\ found that the CN-twist CoIn (Fig.~\ref{fig:cyt:coin2} (c)) is the major relaxation funnel while the oop-NH$_2$ CoIn (Fig.~\ref{fig:cyt:coin2} (b)) plays a secondary role. Additionally, we found a rapid $\pi\pi^*\rightarrow n\pi^*$ transition with a time constant of 40 fs. ISC in the enol form was found much slower than in the keto tautomer, since the singlet-triplet gaps were larger. Additionally, SOCs in the enol form were considerably smaller than in the keto due to the protonation of the oxygen atom.

\subsubsection{Final Discussion}\label{sssec:cyt:sum}

Based on the considerable body of dynamical studies on C, the experimentally observed time constants can be assigned as follows.

The imino tautomer decays the fastest according to Nakayama et al.~\cite{Nakayama2013PCCP}. CASSCF dynamics found a barrierless deactivation path involving the torsion of the imino group. The more accurate CASPT2 predicts the same path, lending high credibility to the results concerning the imino form. Thus, at least part of the fastest time constants ($\tau_1$) obtained in the gas phase experiment could be explained by the decay of imino C, while this tautomer should not contribute to $\tau_2$ and $\tau_3$.

The enol form shows the slowest decay. Thus, it is tempting to assign the slowest experimental time constants to this tautomer. Indeed, the experiments of Kosma et al.~\cite{Kosma2009JACS} and Ho et al.~\cite{Ho2011JPCA} give evidence that the slowest transient vanishes for excitation wavelengths above 280 nm. Above this wavelength, the enol tautomer is not excited anymore, explaining the vanishing transient. Additionally, 1-methyl-cytosine~\cite{Ho2011JPCA}---which does not posses the enol form---also does not show a slow transient, reinforcing the idea that the enol tautomer contributes to $\tau_3$. However, as pointed out by Nakayama et al.~\cite{Nakayama2013PCCP}, the decay mechanism found in the dynamical studies is debatable, since CASSCF dynamics predicts decay via the the oop-NH$_2$ CoIn (Fig.~\ref{fig:cyt:coin2}(b)) but static CASPT2 calculations show a rather high barrier for this CoIn. Instead, CASPT2~\cite{Nakayama2013PCCP} favors decay through a CoIn analogue to the C$_6$-puckered CoIn of the keto form (Fig.~\ref{fig:cyt:coin} (a)). Dynamics simulations based on CASPT2 electronic properties would be required to confirm this hypothesis. The simulations on this tautomer also show a fast component which might also contribute to $\tau_1$. Indeed, unpublished experimental work focusing on the transients of fragment ions specific to the enol tautomer found a transient on the 10 fs timescale~\cite{Weinacht2013private}.

Finally, according to all the dynamics simulations performed, it seems clear that the keto form is responsible for the intermediate $\tau_2$ time constant, although it also contributes to $\tau_1$. It is not clear whether the keto form also contributes to the $\tau_3$ transient. Interestingly, even though most dynamical studies predict qualitatively correct time constants, the proposed mechanistic details differ dramatically---illustrating the importance of the choice of the level of theory. The mechanisms proposed are clearly grouped according to the electronic structure method used. The dynamical studies based on CASSCF employing relatively large active spaces~\cite{Gonzalez-vazquez2010CPC,Barbatti2011PCCP,Richter2012JPCL, Nakayama2013PCCP, Mai2013CPC} predicted the semi-planar CoIn to be the most important pathway for ground state relaxation, while those based on OM2/MRCI~\cite{Lan2009JPCB} or ROKS/BLYP~\cite{Doltsinis2008} favour the C$_6$-puckered CoIn. Indeed, high-level static investigations~\cite{Merchan2006JPCB,Blancafort2007PP,Nakayama2013PCCP} also predict the C$_6$-puckered CoIn to be responsible for the fast decay of keto C. Since this CoIn involves changes at the C$_5$ and C$_6$ (recall Fig.~\ref{fig:cyt:coin}(a)), substitution at C$_5$ should modify the time scales. Indeed, experiments carried out with 5-substituted cytosine derivates~\cite{Middleton2009ARPC} show sensitivity of excited-state lifetimes, confirming the role of this CoIn in the deactivation of the keto form.

In addition to the deactivation pathways in the singlet manifold, the studies based on \textsc{Sharc}~\cite{Richter2012JPCL,Mai2013CPC} showed that ISC from the singlet to the triplet states is of ultrafast nature and contributes to $\tau_2$. This illustrates that $\tau_2$ is actually an average of a number of processes, including ground state relaxation via multiple pathways and ISC. Following ISC, the population gets trapped in the lowest triplet state and ISC from $T_1$ to $S_0$ then provides an explanation for the ns transient observed by Nir et al.~\cite{Nir2002CPL}. One should note that the proposed ISC mechanism found in the dynamics~\cite{Richter2012JPCL, Mai2013CPC} differs from the one given by static calculations~\cite{Merchan2005JACS, Gonzalez-luque2010JCTC}: While the static approach predicted the $^1\pi\pi^*$ state to be the precursor of the triplet states, the dynamics showed that instead $^1n\pi^*$ might be the doorway state. Experimentally~\cite{Middleton2009ARPC}, there is evidence that indeed ISC originates from $^1n\pi^*$. The main argument is that the $^1\pi\pi^*$ population decays too quickly to the ground state for ISC to occur from $^1\pi\pi^*$~\cite{Hare2006JPCB}.

\index{Dynamics simulations!Cytosine|)}
\index{Cytosine|)}


\subsection{Thymine}\label{ssec:results:thy}

\index{Thymine|(}
\index{Dynamics simulations!Thymine|(}

Like C, T is a pyrimidine derivative. In Watson-Crick pairs it is found attached to A. In RNA, it is substituted by U, which is structurally similar, but lacks the methyl group in the 5-position. Although T can exist also in either keto (lactam) or enol (lactim) forms, the keto is the most stable one and predominates both in gas phase and solution; therefore, here by T always the keto form is referred to. 

From the dynamical point of view, T is believed to be the nucleobase with the slowest relaxation upon photo-excitation~\cite{Canuel2005JCP, Kang2002JACS}.


\subsubsection{Experimental Observations}\label{sssec:results:thy:exp}

Table~\ref{tab:thy:exp} collects all the experimentally observed gas phase time scales. Up to four different time scales have been reported in T. Several studies~\cite{Kang2002JACS,Ullrich2004PCCP,Canuel2005JCP,Samoylova2008CP} report a time constants of several ps (5-7 ps). In others~\cite{Ullrich2004PCCP,Canuel2005JCP,Samoylova2008CP} additional shorter time constants in the sub-ps region are found in the excited-state decay of T. Kang et al.~\cite{Kang2002JACS} and Samoylova et al.~\cite{Samoylova2008CP} also observed a weak component in the 100 ps~\cite{Kang2002JACS} or ns~\cite{Samoylova2008CP} region. This is supported by He and coworkers~\cite{He2003JPCA,He2004JPCA}, who found a 22 ns decay.

\begin{table}
  \centering
  \caption{Experimentally observed decay times of T\label{tab:thy:exp}. }
  \begin{tabular}{ccccccllc}
    \hline
    \multicolumn{2}{c}{Setup} & \multicolumn{4}{c}{Time constants} &\multicolumn{2}{l}{Reference} &Year \\
    $\lambda_{\mathrm{pump}}$ (nm) &$\lambda_{\mathrm{probe}}$ (nm) & $\tau_1$ (fs) & $\tau_2$ (fs) & $\tau_3$ (ps)& $\tau_4$ (ns) &\\
    \svhline
    267       &$n\times$800   &       &     &6.4  &$>$0.1  &\cite{Kang2002JACS}    &Kang et al.    &2002\\
    250       &200            &$<$50  &490  &6.4  &        &\cite{Ullrich2004PCCP} &Ullrich et al. &2004\\
    267       &2$\times$400   &105    &     &5.12 &        &\cite{Canuel2005JCP}   &Canuel et al.  &2005\\
    250       &220            &       &     &     &22      &\cite{He2003JPCA,He2004JPCA} &He et al.&2003, 2004\\
    250       &220            &100    &     &7    &$>$1        &\cite{Samoylova2008CP} &Samoylova et al.&2008\\
    \hline
  \end{tabular}
\end{table}


\subsubsection{Deactivation Mechanism}\label{sssec:results:thy:deact}

\begin{figure}
  \includegraphics[scale=1]{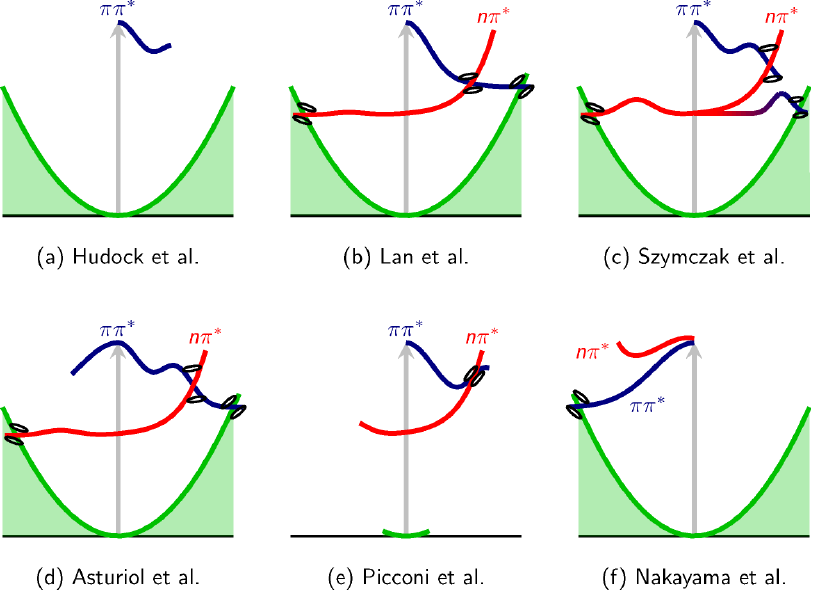}
  \caption{Qualitative overview over the proposed relaxation mechanisms for T from the studies
  \textbf{a}~\cite{Hudock2007JPCA}, 
  \textbf{b}~\cite{Lan2009JPCB},
  \textbf{c}~\cite{Szymczak2009JPCA},
  \textbf{d}~\cite{Asturiol2009JPCA},
  \textbf{e}~\cite{Picconi2011CPC} and
  \textbf{f}~\cite{Nakayama2013JCP}.}
  \label{fig:thy:deact}
\end{figure}

Table~\ref{tab:thy:theo} collects all the excited-state dynamical studies performed in T in the gas phase and Fig.~\ref{fig:thy:deact} summarizes the proposed mechanisms.

\begin{table}
  \centering
  \caption{Overview over excited-state nuclear dynamics studies for isolated T in the gas phase.
  Time constants reported in the papers are given, a checkmark indicates that the authors discussed processes on these timescales without giving explicit values. 
  (classification: $\tau_1$ below 100 fs, $\tau_2$ below 1 ps, $\tau_3$ above 1 ps).
  \label{tab:thy:theo}}
  \begin{tabular}{cccccllc}
    \hline
    \multicolumn{2}{c}{Methodology} &\multicolumn{3}{c}{Time constants}          &\multicolumn{2}{l}{Reference} &Year\\
         Dyn.    &El. Struct.       &$\tau_1$ (fs) &$\tau_2$ (fs) &$\tau_3$ (ps) &&&\\
    \svhline
    FMS          &CASSCF(8,6)       &$\surd$&       &       &\cite{Hudock2007JPCA}       &Hudock et al.  &2007\\
    TSH          &OM2/MRCI          &17     &420    &       &\cite{Lan2009JPCB}          &Lan et al.     &2009\\
    TSH          &CASSCF(10,8)      &100    &       &2.6    &\cite{Szymczak2009JPCA, Barbatti2010PNAS}     &Szymczak et al.&2009\\
                 &                  &       &       &       &                            &Barbatti et al.&2010\\
    TSH          &CASSCF(8,6)       &$\surd$&       &$\surd$&\cite{Asturiol2009JPCA}     &Asturiol et al.&2009\\
    vMCG$^b$     &CASSCF(8,6)       &       &       &       &\cite{Asturiol2010PCCP}     &Asturiol et al.&2010\\
    TSH          &AM1/CI(2,2)       &       &       &       &\cite{Alexandrova2010JPCB}  &Alexandrova et al.&2010\\
    3d-QD        &TDDFT/PBE0        &$\surd$&       &       &\cite{Picconi2011CPC}       &Picconi et al. &2011\\
    MCTDH,       &HLVC model$^a$    &$\surd$&       &       &\cite{Picconi2011CPC}       &Picconi et al. &2011\\
    TSH          &CASPT2(2,2)       &       &400    &       &\cite{Nakayama2013JCP}      &Nakayama et al.&2013\\
    \hline
  \end{tabular}

  $^a$ Harmonic linear vibronic coupling
  $^b$ Variational multi-configurational Gaussians (an MCTDH variant)
\end{table}

\begin{figure}
  \sidecaption
  \includegraphics[scale=1]{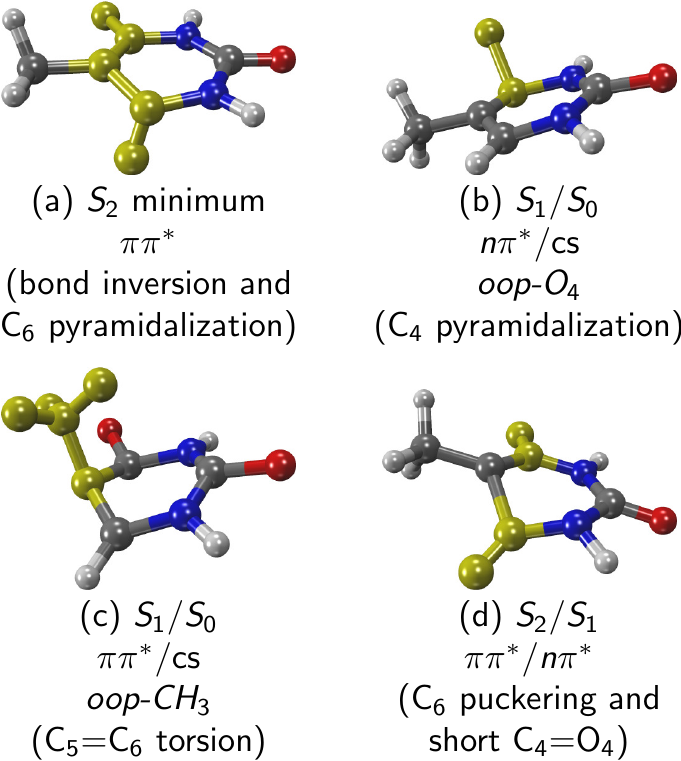}
  \caption{Geometries of important CoIns and minima of T discussed in the text.
  The labels give the crossing states (adiabatic and state character) and the main geometrical feature.}
  \label{fig:thy:coin}
  \index{Conical Intersections!Thymine}
\end{figure}

The first dynamics simulation on T was reported by Hudock et al.~\cite{Hudock2007JPCA} in 2007. Their simulations employed the FMS methodology, coupled to CASSCF(8,6) calculations. This is one of the few papers where an effort is made to make an explicit connection to the experimental results by simulating a time-resolved photoelectron spectrum.The authors reported that during the first 500 fs the system relaxes from the FC region to an $S_2$ minimum. Based on their simulations the authors stated that the shortest time constant in the experiment ($\tau_1$) might be connected to the time the system takes to arrive to the minimum. Even though the study did not extend beyond 500 fs simulation time, they suggested that the experimentally observed ps decay might be related to $S_2\rightarrow S_1$ transfer. The $S_2$ minimum was characterized by a stretching of the C$_5$=C$_6$ and C$_4$=O$_4$ bonds, relative to the FC geometry, and by pyramidalization of the C$_6$ atom (see Fig.~\ref{fig:intro:molecules} for atom numbering). The geometry of the minimum is qualitatively depicted in Fig.~\ref{fig:thy:coin} (a). A summary of the reported pathways is given schematically in Fig.~\ref{fig:thy:deact} (a).

In their 2009 study based on TSH and the semi-empirical OM2/MRCI method, Lan et al.~\cite{Lan2009JPCB} observed a very fast $S_2\rightarrow S_1$ decay (17 fs) and a slower $S_1\rightarrow S_0$ one (420 fs), which correspond to $\tau_1$ and $\tau_2$, respectively. The slower time constant fits well with the 490 fs time constant of~\cite{Ullrich2004PCCP}. No explanation was offered, however, for the 5-7 ps time constants observed in many of the experimental studies~\cite{Kang2002JACS,Canuel2005JCP,Samoylova2008CP}. The two relaxation paths are described as follows, see Fig.~\ref{fig:thy:deact}(b). The fast relaxation of $S_2$ ($\pi\pi^*$) through a CoIn to $S_1$ operates via a planar geometry, which differs only slightly from the $S_2$ minimum (Figu.~\ref{fig:thy:coin}a) with a bond inversion of the C$_6$=C$_5$-C$_4$=O$_4$ moeity. The second and slower relaxation step is mediated by two different CoIns. In the major path, after the hop to $S_1$, the system changes to $n\pi^*$ character, which induces strong out-of-plane motion of the O$_4$ atom. This motion led to the oop-O$_4$ CoIn (Fig.~\ref{fig:thy:coin} (b)), which connects the $n\pi^*$ state with the ground state. A smaller number of trajectories (19\%) relaxed via the direct $\pi\pi^*\rightarrow GS$ path, without intermediate change to $n\pi^*$ character. These trajectories decay via another CoIn, characterized by a strong out-of-plane displacement of the methyl group (Fig.~\ref{fig:thy:coin} (c)).

In the same year, Szymczak et al.~\cite{Szymczak2009JPCA,Barbatti2010PNAS} performed TSH simulations based on CASSCF(10,8). As in the study of Hudock et al.~\cite{Hudock2007JPCA}, here it is found that the slow relaxation of T is primarily related to the trapping of the system in the minimum of the $S_2$ state ($\pi_\mathrm{O}\pi^*$ character). Two time constants were reported~\cite{Szymczak2009JPCA}: $\tau_1=$100 fs, attributed to the initial fast relaxation from the FC region to the $S_2$ minimum and $\tau_2=$2.6 ps for the $S_2\rightarrow S_1$ decay. Also in agreement with Hudock et al.~\cite{Hudock2007JPCA}, they described the $S_2$ minimum geometry with bond inversion of the C$_6$=C$_5$--C$_4$=O$_4$ moeity (Fig.~\ref{fig:thy:deact} (a)). Because the barrier separating the $S_2$ minimum and the $S_2/S_1$ CoIn is rather high at the CASSCF level of theory, IC to $S_1$ is slow. Only after the system surmounts the barrier, IC to $S_1$ quickly proceeds. The $S_2/S_1$ CoIn (crossing of $\pi\pi^*$ and $n\pi^*$) is characterized by a shortening of the C$_4$=O$_4$ bond and puckering of C$_6$ (Fig.~\ref{fig:thy:coin} (d)). After the IC process, the system further relaxes towards the $n\pi^*$ region of the PEH, and is possibly trapped again before ground state relaxation can take place. Consequently, only a small number of trajectories relaxed to the ground state within 3 ps. The additional slow step $S_1\rightarrow S_0$ could explain the discrepancy between the simulated time constant of 2.6 ps and the experimental 5-7 ps time constants. Szymczak et al.\ reported that ground state relaxation could involve either the $\pi\pi^*/S_0$ and $n\pi^*/S_0$ crossing regions, but they did not discuss the geometries involved, so that a connection to the geometries discussed above is not possible. The deactivation paths proposed by Szymczak et al.~\cite{Szymczak2009JPCA} is summarized in Fig.~\ref{fig:thy:deact} (c): trapping in the $\pi\pi^*$ minimum, transfer to $n\pi^*$ and ground state relaxation involving high barriers.

Asturiol et al.~\cite{Asturiol2009JPCA,Asturiol2010PCCP} found two deactivation paths, based on their TSH dynamics employing CASSCF energies. Since at the CASSCF level of theory the $S_2$ $\pi\pi^*$ minimum is separated from the CoIn by a barrier (which is not present at CASPT2 level~\cite{Asturiol2009JPCA}), they started the trajectories instead at the transition state between the minimum and the CoIn. Hence, their simulation does not include the initial trapping observed by Hudock et al.~\cite{Hudock2007JPCA} and Szymczak et al.~\cite{Szymczak2009JPCA}. After going through the $S_2/S_1$ CoIn within 60 fs, the ensemble splits up, with some population continuing towards the $\pi\pi^*/S_0$ CoIn (oop-CH$_3$, Fig.~\ref{fig:thy:coin} (c)), and some going to the $n\pi^*$ state, which finally reach the $n\pi^*/S_0$ CoIn after some time. The latter CoIn was characterized by a long C$_4$=O$_4$ bond, but no pyramidalization of the C$_4$ atom was mentioned. This results are shown schematically in Fig.~\ref{fig:thy:deact} (d). They also noted that at CASPT2 level of theory, there should be an additional, direct path from the FC region to the oop-CH$_3$ CoIn (without trapping in the $S_2$ minimum), which is not well described with CASSCF. On these grounds they proposed that the fast decay components ($\tau_1$ or $\tau_2$) are due to the direct path and the slow component ($\tau_3$) is due to the indirect path which involves $S_2$ trapping.

The second study by Asturiol et al.~\cite{Asturiol2010PCCP} strives to characterize the $S_1/S_2$ CoIn seam in order to investigate the efficiency of entering the $n\pi^*$ state coming from the $\pi\pi^*$ state. Using the variational multi-configurational Gaussians method in 39 DOFs (also based on CASSCF energies), they analyzed the intersection seam of $S_1$ and $S_2$ at regions of different topology. There is a region where the CoIn shows a peaked topology and the $\pi\pi^*$ character is preserved when passing the CoIn, and there is a region with a sloped topology, which allows population transfer to $n\pi^*$. Depending on the initial momentum when approaching the seam, either region can be entered, leading to different photophysical products.

In 2010, another TSH study based on AM1/CI(2,2) semiempirical calculations was published by Alexandrova et al.~\cite{Alexandrova2010JPCB}. As with C, no time constants for isolated T were provided as they focused primarily on nucleosides. For thymidine, the times constants calculated are between 30 and 110 fs. They reported that the relaxation mechanism of T is dominated by ring puckering paths, with puckering at C$_4$ being the most important mechanism.

The work by Picconi et al.~\cite{Picconi2011CPC} focused exclusively on the $\pi\pi^*\rightarrow n\pi^*$ relaxation path (see Fig.~\ref{fig:thy:deact} (e)). They conducted two types of simulations, the first one being a QD study in three dimensions, with PEHs based on TD-DFT with the PBE0 functional. Within this reduced-dimensionality model, the $n\pi^*$ state was populated already within the first 50 fs, in contrast to the findings of Hudock et al.~\cite{Hudock2007JPCA}, Szymczak et al.~\cite{Szymczak2009JPCA} and Asturiol et al.~\cite{Asturiol2009JPCA}. The second simulation of Picconi et al.~\cite{Picconi2011CPC} employed the MCTDH method. The PEHs were calculated with the harmonic linear vibronic coupling model~\cite{Koppel1984}, fitted to TD-DFT energies. These calculations agreed with the 50 fs transfer to the $n\pi^*$ state found in the QD simulations. The authors explained the faster $\pi\pi^*\rightarrow n\pi^*$ transfer as compared to the CASSCF-based dynamical studies with the smaller $n\pi^*-\pi\pi^*$ energy gap predicted by TD-DFT.

The most recent dynamics study on T has been performed by Nakayama et al.~\cite{Nakayama2013JCP} in 2013. In this paper, they reported the first dynamics based on accurate CASPT2 energies and gradients for a nucleobase, although with a small CAS(2,2) active space. To assess the quality of the CASPT2(2,2) calculations, every 50 fs the energies at the current geometry were recalculated with CASPT2(12,9). These calculations showed that the $n\pi^*$ state is above the $\pi\pi^*$ state throughout the simulation. Moreover, NACs were not included in the simulation, so the authors were only able to follow the dynamics on the $\pi\pi^*$ state until the system reached the $\pi\pi^*/S_0$ interaction region, without the possibility to actually hop to the ground state surface.
Interestingly, these calculations showed that in the gas phase the system is not trapped in the $\pi\pi^*$ state, since at CASPT2 level the $\pi\pi^*$ state does not exhibit a minimum.  Instead, (see Fig.~\ref{fig:thy:deact} (f)) the trajectories reach directly the  $\pi\pi^*/S_0$ CoIn (oop-CH$_3$, Fig.~\ref{fig:thy:coin} (c)) after an average of 400 fs. Because the calculations did not include NACs, this time should be considered only a lower bound on the decay time constant (the trajectories could ``miss'' a hop and stay on the $S_1$ PEH for longer time). It should also be noted, that because the simulation included only 10 trajectories and the active space is not flexible enough, it is uncertain whether additional CoIns could be involved in the excited-state dynamics of T.

\subsubsection{Final Discussion}\label{sssec:thy:sum}

In T, much of the theoretical effort involving dynamics simulations has been devoted to study the interplay of the excited $\pi\pi^*$ and $n\pi^*$ states.
The available studies can be classified in two groups, depending whether the $S_2$ minimum is involved or not. The first group includes all CASSCF-based dynamics studies~\cite{Hudock2007JPCA, Szymczak2009JPCA, Asturiol2009JPCA}, where trajectories quickly relax to the $S_2$ $\pi\pi^*$ minimum and get trapped for a considerable amount of time. After leaving the minimum well, the trajectories proceed to the $n\pi^*$ state, from where they eventually decay to the ground state~\cite{Szymczak2009JPCA}. In the second group~\cite{Lan2009JPCB, Picconi2011CPC, Nakayama2013JCP}, photo-excited T does not get trapped in the $\pi\pi^*$ minimum. However, none of the latter dynamics simulations provide an explanation for the experimentally observed 5-7 ps time constant ($\tau_3$). By combining the findings of dynamics and CASPT2 static calculations, Asturiol et al.~\cite{Asturiol2009JPCA} proposed that the biexponential decay could be explained by the bifurcation of the ensemble into a fast, direct relaxation path on the $\pi\pi^*$ state towards the oop-CH$_3$ CoIn and a slower, indirect path involving the $n\pi^*$ state.

A conclusive description of the excited-state dynamics thus might necessitate large-scale dynamics simulations based on highly accurate PEHs, e.g.\ from CASPT2 or MRCI calculations. The study by Nakayama et al.~\cite{Nakayama2013JCP} is already a step in this direction, but the low number of trajectories and the restriction to two states and a small active space did not allow for a complete elucidation of the dynamics of T.

\index{Thymine|)}
\index{Dynamics simulations!Thymine|)}


\subsection{Uracil}\label{ssec:results:ura}

\index{Uracil|(}
\index{Dynamics simulations!Uracil|(}

U is found in RNA exclusively. It is a pyrimidine derivative that forms Watson-Crick pairs with A. In DNA it is replaced by the closely related 5-methyl-U, or T. From the 13 different tautomers possible in U, the di-keto-tautomer is the dominant form in gas phase and solution~\cite{Rejnek2005PCCP} and hence only this form will be discussed henceforth.

\subsubsection{Experimental observations}\label{sssec:results:ura:exp}

Table~\ref{tab:ura:exp} collects experimental studies dealing with the relaxation of photoexcited U and the obtained decay times in gas phase. While the early experiments of Kang et al.~\cite{Kang2002JACS} only found a monoexponential decay of the excited population with a time constant of 2.4 ps, additional time constants were obtained later with the advent of better time resolution. In particular and very similar to the other nuclebases a very short transient ($\tau_1$) is found with constants between 50 and 130 fs. The study of Ullrich et al.~ \cite{Ullrich2004PCCP} reported three transients for U; however, most experiments agree on a biexponential decay behaviour. The recent experiments of Kotur et al.~\cite{Kotur2012IJSTQE} and Matsika et al.~\cite{Matsika2013JPCA} combine TOF-MS with strong field ionization and thus can obtain insights into the differences of U and its fragments, most importantly the fragment with a mass/charge ratio of 69. By comparing the parent ion signal and the $m/Z=$69 signal, they are able to disentangle different decay mechanisms. 

\begin{table}
  \centering
  \caption{Experimentally observed decay times of U\label{tab:ura:exp}. }
  \begin{tabular}{lcccccllc}
    \hline
    Remark&\multicolumn{2}{c}{Setup}& \multicolumn{3}{c}{Time constants} &\multicolumn{2}{l}{Reference} &Year \\
    &$\lambda_{\mathrm{pump}}$ (nm) &$\lambda_{\mathrm{probe}}$ (nm) & $\tau_1$ (fs) & $\tau_2$ (fs) & $\tau_3$ (ps) &\\
    \svhline
          & 267 & n$\times$800 &      &    & 2.4  &\cite{Kang2002JACS}    &Kang et al.    &2002 \\
          & 250 & 200          &$<$50 &530 & 2.4  &\cite{Ullrich2004PCCP} &Ullrich et al. &2004 \\
          & 267 & 2$\times$400 &130   &    & 1.05 &\cite{Canuel2005JCP}   &Canuel et al.  &2005 \\
          & 262 & n$\times$780 &70    &    & 2.15 &\cite{Kotur2012IJSTQE} &Kotur et al.   &2012 \\
    m/z=69& 262 & n$\times$780 &90    &    & 3.21 &\cite{Kotur2012IJSTQE} &Kotur et al.   &2012 \\
          & 262 & n$\times$780 &70    &    & 2.4  &\cite{Matsika2013JPCA} &Matsika et al. &2013 \\
    m/z=69& 262 & n$\times$780 &90    &    & 2.6  &\cite{Matsika2013JPCA} &Matsika et al. &2013 \\
    \hline
  \end{tabular}
\end{table}

\subsubsection{Deactivation mechanism}\label{sssec:results:ura:deact}

\begin{figure}
  \includegraphics[scale=1]{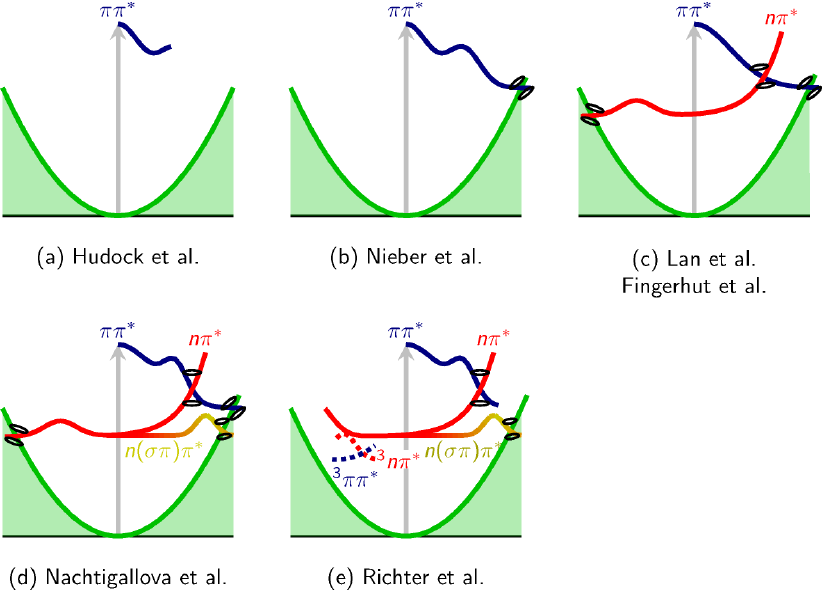}
  \caption{Schematic overview of the proposed relaxation mechanisms for U from
  \textbf{a}~\cite{Hudock2007JPCA}, 
  \textbf{b}~\cite{Nieber2008CP}, 
  \textbf{c}~\cite{Lan2009JPCB, Fingerhut2013JPCL}, 
  \textbf{d}~\cite{Nachtigallova2011JPCA} and
  \textbf{e}~\cite{Richter2014unpublished}.
  Note that different reaction coordinates can be implied in the one-dimensional picture.}
  \label{fig:ura:deact}
\end{figure}

Table~\ref{tab:ura:theo} collects all the excited-state dynamical studies performed in U in the gas phase and Fig.~\ref{fig:ura:deact} summarizes the proposed mechanisms.

\begin{table}
  \centering
  \caption{Excited-state nuclear dynamics studies for isolated U in the gas phase. Time constants correspond to those given in the respective papers, classified as $\tau_1$ below 100 fs, $\tau_2$ below 1 ps, and $\tau_3$ above 1 ps. A checkmark indicates that the authors discussed processes on these timescales without giving explicit values.
  \label{tab:ura:theo}}
  \begin{tabular}{cccccllc}
    \hline
    \multicolumn{2}{c}{Methodology} &\multicolumn{3}{c}{Time constants}          &\multicolumn{2}{l}{Reference} &Year\\
    Dyn.    &El. Struct.            &$\tau_1$ (fs) &$\tau_2$ (fs) &$\tau_3$ (ps) &&&\\
    \svhline
FMS & CASSCF(8,6)         &$\surd$    &    &       &\cite{Hudock2007JPCA}                          &Hudock et al.         &2007 \\
TSH-CP    & ROKS/BLYP &    &551--608 &       &\cite{Nieber2008CP, Doltsinis2008}             &Doltsinis et al.      &2008 \\
                                                                                        &&&&&&Nieber et al.,        &2008 \\
TSH & OM2/MRCI             &21  &570 &       &\cite{Lan2009JPCB}                             &Lan et al.            &2009 \\
TSH & CASSCF(10,8) &    &650--740 &1.5--1.8 &\cite{Barbatti2010PNAS, Nachtigallova2011JPCA} &Barbatti et al.       &2010\\
                                                                                        &&&&&&Nachtigallova et al., &2011 \\
TSH & CASSCF(14,10)        &$\surd$    &$\surd$    &       &\cite{Fingerhut2013JPCL}                       &Fingerhut et al.      &2013 \\
TSH & CASSCF(14,10)        &$\surd$   &    &$\surd$   &\cite{Richter2014unpublished}                                         &Richter et al.        &2014\\
    \hline
  \end{tabular}
\end{table}

The first dynamical study intending to understand the deactivation mechanism of U was done by Hudock et al.~\cite{Hudock2007JPCA} by means of FMS simulations based on CASSCF(8,6) wavefunctions. Very similar to the case of T, after excitation U gets trapped in the $S_2$ minimum (see Fig.~\ref{fig:ura:deact}a). This relaxation pathway is accompanied by an increase of the C$_5$=C$_6$ bond length and a pyramidalization of the C$_6$ atom due to $sp^3$ hybridisation. In their study, they conclude that the ultrafast component $\tau_1$ found in the experiments is neither due to internal conversion via CoIns nor due to a change of character of the excited state. Instead, it is caused by an increase of energy of the ionic states while the system relaxes in the S$_2$, thus making ionization of the excited population less probable. The ps time constant is suggested to be caused by the subsequent barrier crossing from the $S_2$ minimum to a $S_2/S_1$ CoIn and further relaxation. However, the simulation time of 500 fs was too short to see a significant relaxation from the $S_2$ minimum.

\begin{figure}
  \sidecaption
  \includegraphics[scale=1]{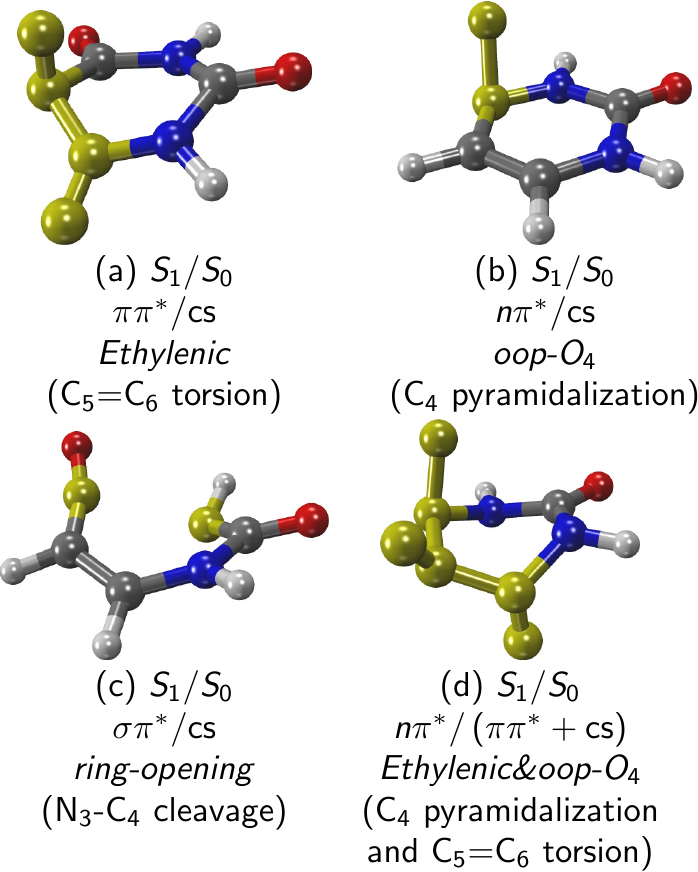}
  \caption{Geometries of important CoIns of U discussed in the text.
  The labels give the crossing states (adiabatic and state character) and the main geometrical feature. }
  \label{fig:ura:coin}
  \index{Conical Intersections!Uracil}
\end{figure}

One year later, Nieber, Doltsinis and coworkers~\cite{Nieber2008CP, Doltsinis2008} reported an study using TSH coupled to Car-Parinello dynamics (TSH-CP) on PEHs calculated with the ROKS/BLYP approach. Their simulations find a sub-ps decay back to the ground state, driven only by $\pi\pi^*/\mathrm{cs}$ interactions and no other electronic states involved (see Fig.~\ref{fig:ura:deact}b). Although they could not reproduce the fast component below 100 fs, they assign this time constant to the initial relaxation from the FC region. They find a monoexponential decay of the excited state population with $\tau_2=$608 fs with non-zero initial velocities ($T$=300 K). Reducing the initial velocities of the trajectories to zero ($T$=0 K) seemed to accelerate the relaxation process to the ground state. Regardless of the initial velocities, the deactivation dynamics is driven by changes in the H--C$_5$=C$_6$--H dihedral angle as well as the C$_5$=C$_6$ and C$_4$--C$_5$ bond length, which give rise to the so-called ethylenic CoIn (see Fig.~\ref{fig:ura:coin}a).

The OM2/MRCI simulations of Lan et al.~\cite{Lan2009JPCB} from 2009 find two different relaxation mechanisms depicted in Fig.~\ref{fig:ura:deact}(c). The first path directly connects the bright $\pi\pi^*$ state with the ground state via the ethylenic CoIn, characterized by a strong out-of-plane motion of the H$_5$ atom, see Fig.~\ref{fig:ura:coin}a. The calculated decay time constant for this path is 21 fs. In the second, slower pathway, the initially excited $\pi\pi^*$ population changes within 70 fs to a lower-lying $n\pi^*$ state via a planar $S_2/S_1$ CoIn that is located close to the FC region. The trajectories spend some time in the $n\pi^*$ state until finally reaching a $S_1/S_0$ CoIn, which was characterized by the authors by a strong out-of-plane motion of the O$_4$ atom (Fig.~\ref{fig:ura:coin}b). For this pathway the decay time constant was reported as 570 fs.

The TSH simulations at the CASSCF(10,8) level of theory, including three states, reported by Barbatti et al.~\cite{ Barbatti2010PNAS} and later in more detail by Nachtigallova et al.~\cite{Nachtigallova2011JPCA} find multiple CoIns and three different deactivation pathways---summarized in Fig.~\ref{fig:ura:deact}(d). The first path agrees with the previous studies: it is found that the fastest decay does not involve a state of another character than the initially excited $\pi\pi^*$ state. After initial trapping in the $S_2$ minimum, the trajectories proceed to a $\pi\pi^*/\mathrm{cs}$ CoIn characterized by a twist of the C$_5$=C$_6$ bond, as shown in Fig.~\ref{fig:ura:coin}a. Interestingly, this path is not observed in a comparable study made on  T~\cite{Szymczak2009JPCA}, even though T and U have very similar PEHs and one could expect a very similar dynamics. The authors~\cite{Barbatti2010PNAS, Nachtigallova2011JPCA} argue that the efficiency of the direct $\pi\pi^*\rightarrow S_0$ path in T is significantly reduced due to the heavy mass of the methyl group. The second pathway found in U is also not observed in T. It involves admixing of $n\pi^*$ character into the wavefunction, weakening the N$_3$--C$_4$ bond to the point of breaking. The involved CoIn, termed ring-opening CoIn and depicted in Fig.~\ref{fig:ura:coin}c, leads to the destabilization of the ground state as the ring breaks. In their study it is stated that the $S_1$ wavefunction contains contributions of $\sigma$ orbitals, which was described as $\sigma(n-\pi)\pi^*$ wavefunction character. The authors also noted that this pathway would probably lead to other photochemical products. Interestingly, Buschhaus et al.~\cite{Buschhaus2013PCCP} indeed observe ring opening after UV irradiation of nucleosides, although the detected isocyanates (R--N=C=O) cannot arise directly from the N$_3$--C$_4$ bond cleavage predicted by Nachtigallova et al.~\cite{Nachtigallova2011JPCA}. The third deactivation pathway described involves a change of character to $n\pi^*$ after initial trapping and relaxation through the $S_2/S_1$ CoIn. The system gets trapped again in the $n\pi^*$ minimum, delaying ground state relaxation. A distinction between $\pi\pi^*$ trapping and $n\pi^*$ trapping was reported based on the significantly different C$_5$--C$_6$ bond lengths of the $\pi\pi^*$ and $n\pi^*$ minima. Two CoIns of $n\pi^*/\mathrm{cs}$ character were discussed. One involves an out-of-plane motion of the O$_4$ atom (see Fig.~\ref{fig:ura:coin}b), as already reported by Lan et al.~\cite{Lan2009JPCB}. The second CoIn~\cite{Barbatti2010PNAS,Nachtigallova2011JPCA} showed a combination of C$_5$=C$_6$ torsion and out-of-plane motion of the O$_4$ atom (see Fig.~\ref{fig:ura:coin}d), resulting in a mixed $n\pi^*$/($\pi\pi^*$+cs) character. Also a dependence of the decay times on the initial energy---corresponding to the excitation wavelengths---was investigated by selecting trajectories with high (250 nm) and low initial energies (267 nm). In both cases, a biexponential decay was observed with time constants of 650 fs and $>$1.5 ps for the high energy and 740 fs and $>$1.8 ps for the low energy trajectories. Here, higher excitation energies reduce the efficiency of trapping in the different excited-state minima, leading to a faster decay to the ground state.

The recent TSH simulations of Fingerhut et al.~\cite{Fingerhut2013JPCL} are based on CASSCF(14,10) wavefunctions. At this level of theory, the two mechanisms described by Lan et al.~\cite{Lan2009JPCB} with semiempirical based dynamics (see Fig.~\ref{fig:ura:deact}c) are found again. First, the initially populated $S_2$ state of $\pi\pi^*$ character decays to the $S_1$ ($n\pi^*$) state, which gains more than 20\% of population within less than 100 fs. In the $n\pi^*$ state, trapping of the population before decay to the ground state can occur, leading to long relaxation times. Interestingly, this study shows much less pronounced trapping in the $S_2$ state than reported by Nachtigallova et al.~\cite{Nachtigallova2011JPCA}, even though both studies use CASSCF. Fingerhut et al.~\cite{Fingerhut2013JPCL} explained this to be an effect of the increased active space size. A second pathway involves the change to $S_1$ without changing the states character and subsequent $\pi\pi^*\rightarrow S_0$ relaxation through the ethylenic CoIn (Fig.~\ref{fig:ura:coin}a). In their simulations, all trajectories that reached the ground state within 1 ps followed the second pathway and only a few trajectories relaxed via the first pathway in longer runs of up to 2 ps.

TSH simulations performed in our group~\cite{Richter2014unpublished} at the same CASSCF(14,10) level of theory but including triplet states show that after 1 ps a significant fraction of the population remains in the $S_2$ ($\pi\pi^*$) state. The relaxation process can be characterized by a biexponential decay. A fast component $\tau_1$ is attributed to the change of state character from the initially excited $\pi\pi^*$. The slower constant $\tau_3$ arises from ISC, directly competing with IC processes. The precursor for ISC was identified as the $S_1$ state, where population is trapped for a sufficiently long time to allow ISC to take place. In contrast to~\cite{Fingerhut2013JPCL},within the first ps of the simulation only a very small number of trajectories relaxed to the ground state, showing that ground state relaxation might be quenched by ISC. Ground state relaxation is mediated by the ring-opening path previously observed by Nachtigallova et al.~\cite{Nachtigallova2011JPCA}.

\subsubsection{Final Discussion}\label{sssec:ura:sum}

As in the other nucleobases, most experimental studies report a very fast time constant $\tau_1$ and two longer time scales ($\tau_2$ and $\tau_3$). Based on the theoretical investigations, the fastest constant $\tau_1$ can be assigned to initial relaxation from the FC region, accompanied by a change of the wavefunction character and an increase in the ionization potential~\cite{Hudock2007JPCA}.

The slower transient ($\tau_2$) has been explained by ground state relaxation, according to dynamics based on ROKS/BLYP~\cite{Doltsinis2008} or OM2/MRCI~\cite{Lan2009JPCB}. CASSCF-based dynamics~\cite{Barbatti2010PNAS, Nachtigallova2011JPCA,Richter2014unpublished} additionally predict even slower processes ($\tau_3$) due to trapping in $S_2$ or $S_1$. The latter studies note that part of the population achieves a fast and direct decay to the ground state---accounting for $\tau_2$---while the remaining fraction of the population causes $\tau_3$. The fact that CASSCF does predict a $\tau_3$ decay---while the other methods do not---emphasizes that the outcome of the dynamics is very sensitive to the electronic structure level of theory.

Due to their very similar molecular structure, T and U exhibit comparable excited-state PEHs and thus share some features of the observed dynamics. Studies reporting dynamics simulations of T and U using the same method~\cite{Hudock2007JPCA,Lan2009JPCB,Barbatti2010PNAS} observe very similar dynamics in both nucleobases. In most of the studies, either the ethylenic CoIn (in T this is the oop-CH$_3$ CoIn) or the oop-O$_4$ CoIn is responsible for the ground state decay. However, it appears that the methyl group which is present in T but not in U might be responsible for the accelerated relaxation dynamics of U. While the ethylenic CoIn in U can be easily reached by torsion of the C$_5$=C$_6$ bond, in T this necessitates the rotation of the bulky methyl group out of the molecular plane.

\index{Dynamics simulations!Uracil|)}
\index{Uracil|)}


\section{Conclusions and Outlook}\label{sec:conclusion}

Upon UV excitation the five nucleobases A, G, C T and U show ultrafast relaxation from the lowest bright $\pi\pi^*$ state to the ground state. This contribution has reviewed the dynamical behaviour that accompanies this process from a dynamical perspective in an effort to interpret the time-resolved spectroscopic experiments from the last decade. While there is an enormous amount of static computations devoted to calculate excite states, potential energy surfaces, and conical intersections of the five molecules, the application of molecular dynamics to nucleobases took off only in 2007. In this chapter we have reviewed 28 publications, which report a total of 35 distinct dynamics simulations. C and T are the bases with the largest number of papers and simulations reported. This fits with the fact that these bases have the slowest and most complex excited-state dynamics. U, which shows a dynamics similar to T has almost as many publications as C or T. The purine bases A and G have a smaller number of publications. For G, only four dynamics simulations have been reported in five papers, but G is also the nucleobases with the fastest and simplest dynamics. The number of simulations for A is relatively large, also taking into account that~\cite{Barbatti2012JCP} contains a comparative investigation using MRCIS, OM2/MRCI and TDDFT with a six different functionals.

\begin{figure}
  \includegraphics[scale=1]{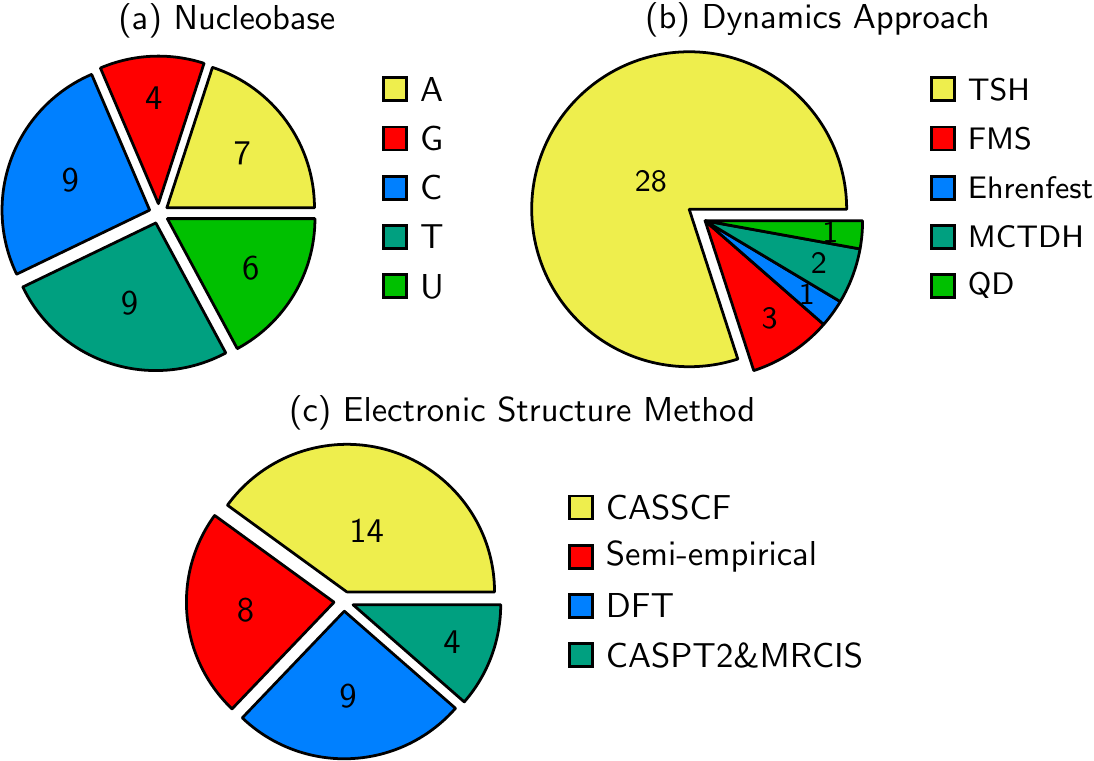}
  \caption{Summary on the number of published nuclear dynamics simulations of isolated nucleobases.
  In \textbf{a}, the number of simulations for each base;
  in \textbf{b} the number of simulations using a given nuclear dynamics approach;
  in \textbf{c} the number of simulations using a given electronic structure method.
  A total of 35 dynamics simulations from 28 publications were considered.}
  \label{fig:concl:piecharts}
\end{figure}

Fig.~\ref{fig:concl:piecharts}a gives the number of simulations performed in each nucleobase. From Fig.~\ref{fig:concl:piecharts}b it is obvious that TSH (including variants like \textsc{Sharc} or TSH-CPMD) is by far the most popular approach for nuclear dynamics simulations, accounting for 80\% of all simulations. In Fig.~\ref{fig:concl:piecharts}c it is possible to appreciate that the most popular electronic structure method underlying dynamics is CASSCF, with 40\% of the simulations, followed by the semi-empirical multi-reference methods and DFT (including TDDFT, ROKS, TDDFTB), each amounting to about 25\%. Only a small number (about 10\%) of the simulations used high-level ab initio methods (MRCIS, CASPT2), in contrast to the large number of static calculations that employ these methods in nucleobases. 

All these calculations show that the ground state relaxation is---in agreement to experimental measurements---happening in all nucleobases within a few ps after excitation. However, the common ultrafast relaxation is not due to a single deactivation mechanism for all nucleobases. There are some general geometrical motifs for the CoIns connecting the excited states to the ground state, like strong puckering of the six-membered rings and bond inversion of C=N--C=O moeities. Nevertheless, the details of the deactivation pathways are very sensitive to the form of the PEHs, which can differ considerably even between tautomers of the same molecule (see e.g.\ Sect.~\ref{ssec:results:cyt} about C).

Generally, the dynamics of the purine bases A and G is less complex and faster than the dynamics encountered in the pyrimidine bases C, T and U. Experiments~\cite{Kang2002JACS, Kang2003JCP, Ullrich2004JACS, Ullrich2004PCCP, Canuel2005JCP} report a biexponential decay for A, with a very fast ($<$100 fs) component and a component of about 1 ps. In G, time constants are even shorter~\cite{Kang2002JACS, Canuel2005JCP}. These findings are consistent with the theoretical dynamical predictions. Based on these simulations, the fast decay can be explained by the fact that the dynamics of A and G is determined by the $\pi\pi^*$ state. The $\pi\pi^*$ state facilitate in both systems a direct route to the CoIn with the ground state, leading to this rapid decay. In A, some studies~\cite{Fabiano2008JPCA, Lei2008JPCA} also predict involvement of the $n\pi^*$ state. Regardless of the underlying electronic structure method, in the purine bases, all CoIns reported in the dynamics studies show a significant degree of ring puckering.

The dynamics of the pyrimidine bases is much slower and significantly more complex. Experimental time constants are around 1-3 ps for C and U and up to 7 ps for T~\cite{Kang2002JACS, Ullrich2004PCCP, Canuel2005JCP}. This slower deactivation has been rationalized by a trapping in an $S_2$ ($\pi\pi^*$) minimum (T, U) or trapping in the $S_1$ ($n\pi^*$) state (C, T, U). The formation of an $\pi\pi^*$ minimum has been attributed to the interaction of different $\pi\pi^*$ states~\cite{Hudock2007JPCA, Szymczak2009JPCA, Barbatti2010PNAS}. It was also found that different $n\pi^*$ states (e.g.\ located at the carbonyl oxygen or nitrogen atoms in the ring) may interact in certain regions of the PEHs. Even though some studies predicted trapping, there are also simulations which show relaxation to the ground state much faster (sub-ps), at least for part of the ensemble. The observed multiple timescales thus could be explained by a splitting of the wavepacket, with one part of the wavepacket decaying directly and another fraction getting trapped.

In general, the presented dynamics simulations still suffer on one side from the limited quality of the underlying electronic structure methods and on the other side from lack of a proper description of the probe process in the simulations. Highly accurate correlated multi-reference methods are still too expensive and/or are not yet technically prepared to be efficiently employed in dynamics calculations of systems of the size of nucleobases. Yet, with the rapid advance of electronic structure codes, an impetuous development in the field dynamics is expected, allowing to deliver quantitative or semi-quantitative results. The simulation of dynamical processes including the actual pump and probe laser pulses---as it is done in the experiment---is yet in its infancy. Progress in this direction will also be witnessed in the future, considerably increasing our understanding on the photophysics and photochemistry of nucleobases.


\begin{acknowledgement}
Financial support from the Austrian Science Fund (FWF), Project No. P25827 is gratefully acknowledged. Furthermore, we would like to thank Jesus Gonz\'alez-V\'azquez and Tom Weinacht for their always insightful discussions. Special thanks goes to Tom for sharing his unpublished results on enol cytosine with us. The Vienna Scientific Cluster (VSC) is also thanked for generous allocation of computer time.
\end{acknowledgement}


\printindex

\end{document}